\documentclass[a4paper,11pt]{article}
\usepackage[top=3cm,right=3cm,bottom=2.5cm,left=2.5cm]{geometry}  
\usepackage[utf8]{inputenc}
\usepackage{setspace}
\usepackage{hyperref}
\usepackage{amsmath}
\usepackage{amsthm}
\usepackage{amssymb}
\usepackage{amsfonts}
\usepackage{dsfont}
\usepackage{mathrsfs}
\usepackage{graphicx} 
\usepackage{cite}
\usepackage{hyperref} 
\usepackage{cleveref}
\newcommand{\ud}{\,\mathrm{d}}

\DeclareMathOperator{\Bigcup}{\bigcup}
\newcommand{\deltabar}{{\mkern0.75mu\mathchar '26\mkern -9.75mu\delta}}

\begin{document}
	\vspace*{1cm}
	\begin{center}
		{\Large{On variational principle and canonical structure of gravitational theory in double-foliation formalism}}
		
		\vspace*{1cm}	
		{\large Sajad Aghapour$^{ab,}$\footnote{aghapour@ipm.ir},  Ghadir Jafari$^{a,}$\footnote{ghjafari@ipm.ir}, Mehdi Golshani$^{ab,}$\footnote{golshani@theory.ipm.ac.ir}}
		\vspace*{1cm}
		
		{\it $^a$ School of Physics, Institute for Research in Fundamental Sciences (IPM), P.O. Box 19395-5531, Tehran, Iran
			
			$^b$ Physics Department, Sharif University of Technology, P.O. Box 11365-8639 , Tehran, Iran
		}
	\end{center}
	\vspace*{1cm}
	
	\begin{abstract}
		In this paper, we analyze the variation of the gravitational action on a bounded region of spacetime whose boundary contains segments with various characters, including null. We develop a systematic approach to decompose the derivative of metric variations into orthogonal and tangential components with respect to the boundary and express them in terms of variations of geometric objects associated with the boundary hypersurface. We suggest that a double-foliation of spacetime provides a natural and useful set-up for treating the general problem and clarifies the assumptions and results in specialized ones. In this set-up, we are able to obtain the boundary action necessary for the variational principle to become well-posed, beside the canonical structure of the theory, while keeping the variations quite general. Especially, we show how one can remove the restrictions imposed on the metric variations in previous works due to the assumption that the boundary character is kept unaltered. As a result, we find that on null boundaries a new canonical pair which is related to the change in character of the boundary. This set-up and the calculation procedure are stated in a way that can be applied to other more generalized theories of gravity.
	\end{abstract}
	
	\section{Introduction}
	Einstein-Hilbert action, which is defined on the bulk of spacetime, contains second order derivatives of the metric, the dynamical field of the theory. As a consequence, the variational principle is not well-posed for this action unless some additional terms, defined on the spacetime  boundary (and its corners; see below), are considered. Therefore, the gravitational action consists of a bulk term and a boundary term (and corner terms as well). When the boundary is time-like or space-like, such a boundary action is given by the well-known Gibbonse-Hawking-York term\cite{Gibbons:1976ue}. In the case of null boundaries, there have been investigations recently by various groups to propose the appropriate boundary action  \cite{Parattu:2015gga,Lehner2016,Hopfmuller2017,Jubb:2016qzt,Neiman:2012fx}. 
	
	Generally, the boundary action can be found directly by varying the action and using the fact that there appear some total variations that are required to be removed so that the variational principle becomes well-posed. The proper boundary action could be read from those total variations. This approach is explained in \cite{Padmanabhan2014} and used to derive the boundary action on null boundaries in \cite{Parattu:2015gga} for the first time. An alternative approach includes topological considerations where one seeks the boundary terms necessary for constructing the correct Euler character in the case of bounded manifolds \cite{Myers1987}. The latter approach, while interesting, is only applicable for Lanczos-Lovelock theories of gravity, not to mention that it has not been used in the case of null boundaries yet due to difficulties.
	The variational approach has also this advantage that it gives, beside the boundary action, the canonical structure of the theory and physical degrees of freedom (see for example \cite{misner2017gravitation,Parattu:2015gga,Hopfmuller2017}), whereas the topological approach does not.
	
	The variation of gravitational action, beside the equations of motion, consists in a surface integral that, in general, is a functional of the metric variation and its derivatives. Explicitly, the Einstein-Hilbert action
	\begin{align}
	\mathcal A_{\scriptscriptstyle{EH}} = \int \ud^d x \sqrt{-g} \,R\,,
	\end{align}
	when is varied, leads to
	\begin{align}
	\delta \mathcal A_{\scriptscriptstyle{EH}} = \int \ud^d x \sqrt{-g} \,(-G^{ab} \delta g_{ab} + g^{ab} \delta R_{ab})
	\end{align}
	in which the first term gives the equations of motion and the second term is a surface integral that can be written as
	\begin{align}
	\Theta_{\scriptscriptstyle\partial V} \equiv \int_V \ud^d x \sqrt{-g} \, g^{ab} \delta R_{ab} &= \int_V \ud^d x \sqrt{-g} \, \nabla_a (g^{bc}\,\delta\Gamma^a_{bc} - g^{ab}\,\delta\Gamma^c_{bc}) \nonumber\\
	&= \int_{\partial V} \ud^{d-1} x \sqrt{-g} \, v_a  (g^{bc}\,\delta\Gamma^a_{bc} - g^{ab}\,\delta\Gamma^c_{bc}) \nonumber\\
	&= 2\int_{\partial V} \ud^{d-1} x \sqrt{-g} \, P^{abcd} \, v_a  \nabla_d\,\delta g_{bc} \label{boundaryInt}
	\end{align}
	where  $ v_a=\nabla_a\phi $ is the outward normal to the boundary, characterized by $ \phi=\text{const.} $, and $P^{abcd} = \frac12(g^{ac}g^{bd}-g^{ad}g^{bc})$
	\footnote{The last expression in \eqref{boundaryInt} is, in fact, the form of boundary term for Lanczos-Lovelock action in which $P^{abcd} = \frac{\delta \mathcal L}{\delta R_{abcd}}$. It is clear the $P^{abcd}$ has the symmetries of Riemann curvature tensor; see for example\cite{Chakraborty2017}.}. It is easy to see that the last expression in \eqref{boundaryInt} contains the metric variation and both of its tangential and normal derivatives. 
	
	Generally speaking, one can disintegrate a surface term like \eqref{boundaryInt} in the variation of action, with the aid of decomposition vis-à-vis the boundary hypersurface and writing different contributions in terms of geometrical objects associated with the boundary, schematically in the form
	\begin{align}
	\Theta_{\scriptscriptstyle\partial V} = \delta(\int_{\mathcal B_i} A_i) + \delta(\int_{\mathcal C_i} a_i) + \int_{\mathcal B_i} P_i\,\delta Q_i + \int_{\mathcal C_i} p_i\,\delta q_i \label{schematicTheta}
	\end{align}
	where $\mathcal B_i$s are boundary segments: $\partial V= \Bigcup_i \mathcal B_i$ and $\mathcal C_i$s are co-dimension two corners, i.e. intersections of neighboring segment $\mathcal B_i$s. 
	The first two sets of terms in \eqref{schematicTheta}, which are total variations, suggest the boundary and corner terms to be included in the action on boundary segments $\mathcal B_i$s and corners $\mathcal C_i$s. As a result, the modified action will contain the bulk, the boundary and the corner terms:
	\begin{align}
	\mathcal A' = \mathcal A_{\scriptscriptstyle{bulk}} - \int_{\mathcal B_i} A_i - \int_{\mathcal C_i} a_i \label{action_general_form}
	\end{align}
	and its variation does not carry the total variation terms so that the variational principal becomes well-posed.
	
	The last two sets of terms in \eqref{schematicTheta} give the canonical structure: the canonical pairs of configuration variables and their conjugate momenta $(Q_i, P_i)$ on boundary segments and $(q_i, p_i)$ on corners. The canonical structure of the theory gives clues to realizing the degrees of freedom. For General Relativity, the canonical pair on non-null hypersurfaces is well-known to be $(h_{ab}, K_{ab} - K\,h_{ab})$, where $h_{ab}$ is the induced metric on the hypersurface and $K_{ab}$ is its extrinsic curvature. The canonical structure of GR on  null hypersurfaces, beside the the boundary action terms, has been examined recently, using variational methods in \cite{Parattu2016,Lehner2016,Hopfmuller2017,Jubb:2016qzt}. 
	
	Here we describe a systematic procedure to find the canonical structure, beside the boundary and corner terms of the action, by decomposing the variations with respect to a frame adapted to the boundary hypersurface. An important point in finding the complete canonical structure is that it is necessary not to suppress any degree of freedom beforehand by imposing restrictions. The criteria that whether a variable is a physical degree of freedom of the theory is that if it enters in the symplectic structure or not. We discuss the usual restrictions on variations, especially those that keep the character of the boundary unchanged, and try to remove these restrictions and study the consequences. In particular, we show that one of the degrees of freedom is responsible for the change in the character of the hypersurface. Therefore, if we don't restrict the variation of metric to preserve the character of the hypersurface, we can obtain an extra degree of freedom entering in the canonical structure which is absent in \cite{Parattu2016,Hopfmuller2017}.
	
	In our calculational set-up, we benefit from a double-foliation of the spacetime. Such foliations of spacetime are proved to be useful in a variety of circumstances in gravitational physics. For example, it is unavoidable to doubly foliate the spacetime when dealing with null hypersurfaces and studying their dynamics (see e.g. \cite{Sachs1962,Hayward1993,Brady1995,Epp1995,Vickers2011}). Usually, the double foliations introduced in these studies are formed by two sets of null hypersurfaces. However, the double-null-foliation is not adequate for the cases where spacetime dynamics permits the evolution of null surfaces to become non-null. Following \cite{Hopfmuller2017}, we use a framework where the double foliation is chosen to be general, i.e. the leaves of foliation are allowed to be of any character. This choice has the advantage of being free of any gauge-fixing beforehand. We show that double-foliation is also useful in non-null boundary cases.
	
	The rest of the paper is organized as follows: In section \ref{non-null case}, we review the known results for time-like and space-like boundaries, where we explain the procedure that we use for decomposition of metric variation and its derivative, as well as the geometrical objects associated with the boundary, to calculate the boundary and corner terms needed for modified action. Based on this procedure, in section \ref{nul case}, we consider the more tough case of null boundary segments and derive the canonical structure and boundary action in a systematic manner and find the canonical pair related to the change of the boundary character. In section \ref{general case}, we generalize the setup in order to consider all types of boundaries in a unified treatment.
	
	\section{Non-null Boundary Segments}\label{non-null case}
	Suppose that the boundary segment $\mathcal B$ under consideration is non-null i.e. it is time-like or space-like. Let this boundary be a level surface of a scalar field $\phi$. We define the 1-form $n_a$ through the spacetime adapted so that on the boundary, $n_a = N\,\nabla_a\phi = N\,v_a$. We also require $n_a$ to be normalized everywhere to $\epsilon$ ($\epsilon=+1$ in the case of time-like boundary and $\epsilon=-1$ for space-like one).
	
	The boundary term \eqref{boundaryInt} can be recast as
	\begin{align}
	\Theta_{\scriptscriptstyle\mathcal B}= 2\int_{\mathcal B} \ud^{d-1} x \sqrt{|h|} \, P^{abcd} \, n_a  \nabla_d\,\delta g_{bc}
	\end{align}
	where $h$ is the determinant of the induced metric on $\mathcal B$ and we have used $\sqrt{-g} = N\sqrt{|h|}$.
	
	\subsection{Decomposition with respect to the boundary}
	With the aid of unit normal 1-form $n_a$, we can define the projector $h^a{}_b = \delta^a{}_b - \epsilon\, n^a\,n_b$ to be able to decompose tensors perpendicular and tangent to $\mathcal B$. For example, $\nabla_a n_b$, which we encounter frequently, can be decomposed as:
	\begin{align}
	\nabla_a \, n_b &= \delta^c{}_a\,\delta^d{}_b\,\nabla_c \, n_d = (h^c{}_a + \epsilon\, n^c\,n_a)(h^d{}_b + \epsilon\, n^d\,n_b)\nabla_c \, n_d \nonumber\\
	&= h^c{}_a\,h^d{}_b\,\nabla_c \, n_d + \epsilon\, n^c\,n_a\, h^d{}_b \,\nabla_c\, n_d = -K_{ab} + \epsilon\,n_a\,a_b \label{delndecompose}
	\end{align}
	where we have used $n^a\,\nabla_b n_a = \tfrac12\nabla n^2 = 0$ and the definitions of extrinsic curvature $K_{ab}$ of the boundary and the acceleration of its normal are:
	\begin{align}
	&K_{ab} = -h^c{}_a\,h^d{}_b\,\nabla_c \, n_d \label{Kdef} \\
	&a_b=  h^d{}_b \,n^c\,\nabla_c\, n_d \label{adef}
	\end{align}
	
	\subsection{Variations and their decomposition}
	The procedure that we use to analyze the boundary term is as follows: first we decompose the variation of spacetime metric $\delta g_{ab}$ with respect to the boundary, and define
	\begin{align}
	&\deltabar  h_{ab} = h^c{}_a\, h^d{}_b\,\delta g_{cd}\,,\\
	&\deltabar  u_a = \epsilon\, h^b{}_a \, n^c\,\delta g_{bc} =\epsilon\, n^b \, h^c{}_a \,\delta g_{bc}\,,\\
	&\deltabar \mu = n^a \, n^b\,\delta g_{ab}\,,
	\end{align}
	so that
	\begin{align}
	\delta g_{ab} = \deltabar \mathit{h}_{ab} + 2\,\deltabar \mathit{u}_{(a}\, \mathit{n}_{b)}  + \deltabar \mu\, \mathit{n}_{a} \mathit{n}_{b}\,. \label{deltagdecompose}
	\end{align}
	Note that in our notations, $\deltabar  h$, $\deltabar u$ and $\deltabar \mu$ are not variations of any functions, especially $\deltabar h_{ab} \ne \delta h_{ab}$, but $\deltabar h_{ab} = h^c{}_a\,h^d{}_b\,\delta h_{cd}$.
	
	Then we consider the variation of $n_a$. For $n_a = N \nabla_a\phi$, with the assumption\footnote{This assumption seems quite natural, since one expects that the description of the hypersurface is unchanged under the variations of metric.} $\delta\phi=0$, and therefore $\delta\nabla_a\phi=0$,
	we obtain
	\begin{align}
	\delta n_a = (\delta N)\,\nabla_a\phi = \frac{\delta N}N \, n_a = (\delta\ln N) \, n_a \label{deltan2}
	\end{align}
	On the other hand, the normalization condition of $n_a$ makes a relation between its variation and variation of the spacetime metric $g_{ab}$, which is assumed to be the physical field. Therefore we have
	\begin{align}
	0=\delta(n^a\,n_a)= 2\,n^a\,\delta n_a - n^a\,n^b\,\delta g_{ab} = 2\,n^a\,\delta n_a - \deltabar \mu \label{deltan^2=0}
	\end{align}
	Comparing \cref{deltan2,deltan^2=0}, we find that
	\begin{align}
	\deltabar \mu = 2\,\epsilon\,\delta(\ln N) \,. \label{deltamu}
	\end{align}
	Note that in this set-up the character of the boundary cannot change; namely a space-like boundary remains space-like and so on. This can be seen by considering that under the metric variations, if we have $n_a \to n'_a = n_a + \delta n_a$, then 
	$n'^a\,n'_a = N'^2 (\nabla^a\phi\nabla_a\phi)'$. As a result, for example, the variation of a space-like surface to a null surface yields $(\nabla^a\phi\nabla_a\phi)'=0$, which indicates that $ N'^2$ diverges and the set-up fails. Similarly, for a variation from a space-like to a time-like surface, there is no smooth variation.
	One can avoid this problem by using $v_a = \nabla_a \phi$ directly, instead of the normalized $n_a$. An alternative approach is to use a double-foliation of spacetime. This latter approach provides a deeper understanding of the problem in connection with the situation we encounter in the case of null boundary segments. We will discuss this further  in section \ref{general case}.

	\subsection{Decomposition of $\nabla_a \delta g_{bc}$}
	Now we are at the position to decompose the expression $\nabla_a \delta g_{bc}$, appearing in relation \eqref{boundaryInt}. Taking a covariant derivative of \eqref{deltagdecompose}, we find
	\begin{align}
	\nabla_a (\delta g_{bc}) &= \tfrac12\,\nabla_a (\deltabar h_{bc}) + n_c\,\nabla_a (\deltabar u_{b}) + \tfrac12\, n_b\,n_c\,\nabla_a(\deltabar \mu) +( \deltabar u_b + \deltabar \mu\,n_b)\,\nabla_a n_c \nonumber\\
	& \quad + (b\leftrightarrow c)\,, \label{deldeltagdecompose}
	\end{align}
	that we need to decompose further with respect to the boundary. For the first term in \eqref{deldeltagdecompose} we have
	\begin{align}
	\nabla_a (\deltabar h_{bc}) &= (h^d{}_a - \epsilon\,n^d\,n_a)(h^e{}_b - \epsilon\,n^e\,n_b)(h^f{}_c - \epsilon\,n^f\,n_c)\,\nabla_d(\deltabar h_{ef}) \nonumber\\
	&= D_a (\deltabar h_{bc}) + \epsilon\,n_a\, h^e{}_b\,h^f{}_c\,n^d\, \nabla_d (\deltabar h_{ef}) - 2\,\epsilon\,  n_{(b}\,\deltabar h_{c)d}\,\nabla_a n^d \label{deldeltag1st}
	\end{align}
	where $D_a (\deltabar h_{bc})=h^d{}_a\,h^e{}_b\,h^f{}_c\, \nabla_d (\deltabar h_{ef})$, $D_a$ being the intrinsic covariant derivate of the boundary and we have used $n^b \, \nabla_a (\deltabar h_{bc}) = - \deltabar h_{bc} \nabla_a n^b$, which results from $n^a \, \deltabar h_{ab} =0$.
	\\ \noindent
	For the second term of \eqref{deldeltagdecompose}, we have
	\begin{align}
	\nabla_a(\deltabar u_b) &= (h^c{}_a - \epsilon\,n^c\,n_a)(h^d{}_b - \epsilon\,n^d\,n_b)\,\nabla_c(\deltabar u_d) \nonumber\\
	&= D_a(\deltabar u_b) + \epsilon\,n_a\,h^d{}_b\,n^c\,\nabla_c(\deltabar u_d) -\epsilon\, n_b\,\deltabar u_c \,\nabla_a n^c \label{deldeltag2nd}
	\end{align}
	where we have used $n^b \, \nabla_a (\deltabar u_{b}) = - \deltabar u_b \nabla_a n^b$, which results from $n^a \, \deltabar u_{a} =0$.
	\\ \noindent
	Finally, the third term of \eqref{deldeltagdecompose} yields
	\begin{align}
	\nabla_a(\deltabar \mu) = D_a(\deltabar \mu) -\epsilon\,n_a\,n^b\,\nabla_b(\deltabar \mu) \label{deldeltag3rd}
	\end{align}
	Replacing \eqref{deldeltag1st}, \eqref{deldeltag2nd} and \eqref{deldeltag3rd} in \eqref{deldeltagdecompose} and using the decomposition of $\nabla_a n_b$ in \eqref{delndecompose}, we obtain
	\begin{align}
	\nabla_a (\delta g_{bc}) &= \tfrac12\,D_a (\deltabar h_{bc}) + \tfrac12\epsilon\,n_a\, h^e{}_b\,h^f{}_c\,n^d\, \nabla_d (\deltabar h_{ef}) \nonumber\\
	&\quad+ n_c\,D_a(\deltabar u_b) + \epsilon\,n_a\,h^d{}_b\,n_c\,n^e\,\nabla_e(\deltabar u_d) \nonumber\\
	&\quad+ \tfrac12\, n_b\,n_c\,D_a(\deltabar \mu) -\tfrac12\, \epsilon\,n_a\,n_b\,n_c\,\,n^d\,\nabla_d(\deltabar \mu)\nonumber\\
	&\quad+(-\epsilon\,n_b\,\deltabar h_{ce} + h_{be}\,\deltabar u_c - \epsilon\, n_b\,n_c\,\deltabar u_e + h_{be}\,n_c\,\deltabar \mu)\nabla_a n^e \nonumber\\
	& \quad + (b\leftrightarrow c) \label{deldeltagdecompose2}
	\end{align}
	
	The expression \eqref{deldeltagdecompose2} can be written in a more useful form in terms of $\delta K_{ab}$ and $\delta a_a$. To this end, we compute the variation of $K_{ab}$ and $a_a$ in \cref{Kdef,adef} and decompose them to find:
	\begin{align}
	\delta K_{ab} &= \tfrac12(K_a{}^c\,\deltabar h_{bc} - a_a\,\deltabar u_b + n_a \,K_{bc}\,\deltabar u^c - \tfrac12\,\epsilon\,K_{ab}\,\deltabar \mu  \nonumber\\
	&\quad - \tfrac12\,h^d{}_a\,h^e{}_b\,n^c\,\nabla_c(\deltabar h_{de}) + \epsilon\,D_a(\deltabar u_b) )+ (a\leftrightarrow b) \label{deltaK}\\
	\delta a_a &= n_a\,a^b\,\deltabar u_b -\tfrac12\, D_a(\deltabar \mu)\,. \label{deltaa2}
	\end{align}
	Then, we solve \eqref{deltaK} for $\tfrac12\,h^d{}_a\,h^e{}_b\,n^c\,\nabla_c\deltabar h_{de}$  and \eqref{deltaa2} for $D_a(\deltabar \mu)$, and substitute them in \eqref{deldeltagdecompose2} to obtain the final result for $\nabla_a(\delta g_{ab})$:
	\begin{align}
	\nabla_a (\delta g_{bc}) =& \,-\epsilon\,h^d{}_b\,h^e{}_c\,n_a\,\delta K_{de} - n_b\,n_c\,\delta a_a + (2\,\epsilon\,n_{(a}\,K_{c)}{}^d - n_a\,n_c\,a^d)\deltabar h_{bd}\nonumber\\
	& -(K_{ab}\,g_{cd}  - \epsilon\,K_{ad}\,n_b\,n_c )\deltabar u^d\nonumber\\
	& -(\tfrac12\,n_a\,K_{bc}+K_{ac}\,n_b-\epsilon\,n_a\,a_b\,n_c)\deltabar \mu\nonumber\\
	& +\tfrac12\,D_a(\deltabar h_{bc})+ 2\,D_c(\deltabar u_{(a})\,n_{b)} +\epsilon\,h_{ce}\,n_a\,n_b\,n^d\,\nabla_d(\deltabar u ^e)\nonumber\\
	&  + \tfrac12 \epsilon\,n_a\,n_b\,n_c\,n^d\,\nabla_d(\deltabar \mu) + (b\leftrightarrow c) \,.\label{finalCDG}
	\end{align}
	This explicit expression for $\nabla_a(\delta g_{ab})$ in terms of geometric objects and their variations is appearing here for the first time. It is a very useful expression in deriving the surface term and fining the boundary action of various theories. For example, for Lanczos-Lovelock theory, using the symmetries of $P^{abcd}$, we find
	\begin{align}
	2 \,P^{abcd}\,n_a\,\nabla_d (\delta g_{bc}) = & 4\,\epsilon\,A^{ab}\,\delta K_{ab} - 2\,\epsilon\,A^{ab}\,K_a^c\,\deltabar h_{bc} + 2\,B_a^{bc}\,K_{bc}\,\deltabar u^a \nonumber\\
	&- 2\,B^{abc}\,D_b\,(\deltabar h_{ac}) + 2\,A^{ab}\,D_a(\deltabar u_b) \label{lovelocksurfaceterm}
	\end{align}
	where
	\begin{align*}
	&A_{ab} = h^e{}_a\,n^c\,h^f{}_b\,n^d\,P_{ecfd} \ ,\quad B_{abc} = h^e{}_a\,h^f{}_b\,h^g{}_c\,n^d\,P_{efgd}
	\end{align*}
	The equation \eqref{lovelocksurfaceterm} coincides with the result of the recent paper \cite{Chakraborty2017}.
	
	\subsection{Disintegration of non-null surface term}
	For Einstein theory, where $P^{abcd} = \frac12(g^{ac}g^{bd}-g^{ad}g^{bc})$, we find
	\begin{align}
	2\,P^{abcd}\,n_a\,\nabla_d (\delta g_{bc}) = & \, 2\,h^{ab}\,\delta K_{ab} - K^{ab}\,\delta h_{ab} - \epsilon\,D^a(\deltabar u_a)\,, \label{nonnullEinsteinBT}
	\end{align}
	in which we have used the fact that $h_a{}^c\,h_b{}^d\,\deltabar h_{cd} = \delta h_{ab}$. Using \eqref{nonnullEinsteinBT}, the surface term \eqref{boundaryInt} becomes:
	\begin{align} \label{BTnonnull1}
	\Theta_{\mathcal B}=
	&\int_{\mathcal B}\!\!\! \ud^{d-1} x \,\frac{\sqrt{-g}}N \left[ 2\,h^{ab}\,\delta K_{ab} - K^{ab}\delta h_{ab} -\epsilon D^a(\deltabar u_a)\right] \nonumber\\
	=&\int_{\mathcal B}\!\!\! \ud^{d-1} x \,\sqrt{|h|} \left[ 2\,\delta(h^{ab}\, K_{ab}) + K^{ab}\,\delta h_{ab}-\epsilon D^a(\deltabar u_a)\right]  \nonumber\\
	=&\int_{\mathcal B}\!\!\! \ud^{d-1} x  \left[ 2\,\delta(\sqrt{|h|}\, K) + (K^{ab}-K\,h^{ab})\,\delta h_{ab}-\epsilon D^a(\deltabar u_a)\right]  
	\end{align}
	The first two terms reside on $\mathcal B$, while the last term can be thrown on the boundary of $\mathcal B$, which is a co-dimension two surface $\partial\mathcal B$. To do so, we consider a foliation within $\mathcal B$ by $\psi$ with $\partial\mathcal B$ as its level surface. Then the unit normal of $\partial\mathcal B$, which is tangent to $\mathcal B$, can be defined as $m_a=M\nabla\psi$. On $\mathcal B$, we have $ \sqrt{-g}=N\sqrt{|h|} $ and on $\partial\mathcal B$, $ \sqrt{|h|}=M\sqrt{|q|} $, where $h$ and $q$ are the determinants of the induced metrics on $\mathcal B$ and $\partial\mathcal B$, respectively. So \eqref{BTnonnull1} can be restated as 
	\begin{align}
	\Theta_{\mathcal B}=&2\delta(\int_{\mathcal B}\!\!\! \ud^{d-1} x \,\sqrt{|h|} K)\nonumber\\&+\int_{\mathcal B}\!\!\! \ud^{d-1} x \sqrt{|h|} (K^{ab}-K h^{ab})\delta h_{ab}\nonumber\\&
	+\int_{\partial\mathcal B}\!\!\! \ud^{d-2} x \sqrt{|q|}\, m^a\,\deltabar u_a\,. \label{BTnonnull}
	\end{align}
	The first term suggests the boundary action that we need in order to have a well-defined variational problem, the well-known Gibbons-Hawking term. The second term gives the canonical pair of the theory. The last term is unsavory, because it is neither in the form of a total variation nor a canonical pair. The known remedy for this problem, as in
	\cite{Lehner2016,Hayward:1993my}, is to add up two such terms from neighboring boundary segments to build a total variation on their intersection. This leads to the so-called corner terms being usually the angle or boost parameter between unit normals of the two boundary segments. We evaluate the corner terms in a similar manner in the next subsection.
	
	At this stage, let us make some remarks on counting the degrees of freedom of gravitational field, when the spacetime has non-null boundaries. To identify the physical degrees of freedom in such a boundary value problem, one needs to realize the boundary data, which have to be fixed on the boundary to generate different solutions of equations of motion, beside the constraints imposed on those data or equivalences among the generated solutions. The canonical structure inferred from equation \eqref{BTnonnull} characterizes the dynamical variables that their value and/or their normal derivatives have to be fixed on the boundary. For example, it is easily seen that the Dirichlet boundary conditions on a non-null boundary $\mathcal B$ is $\delta h_{ab} = 0$. The pair of variables $(h_{ab},K_{ab}-K\,h_{ab})$ are exactly the canonical pair of configuration and its conjugate momentum variables that emerge in the Hamiltonian approach to gravity. The induced metric $h_{ab}$, as a symmetric rank-2 tensor on a 3-surface, has six independent components. The conjugate momentum to $h_{ab}$, which is completely determined by the extrinsic curvature $K_{ab}$, has also six independent components. These are twelve functions on the boundary segment $\mathcal B$; but there are four constraint equations on these functions, namely the nondynamical subset of equations of motion:
	\begin{align}
	&G_{ab}\,n^a\,n^b = \tfrac12(\varepsilon\,{}^3R - K^{ab}\,K_{ab} + K^2)=0\\
	&G_{ab}\,h^a_c\,n^b = D_a(K^{a}{}_c - K\,h^{a}{}_c) =0 \,.
	\end{align}
	These equations are called Hamiltonian and momentum constraints respectively. They effectively reduce the independent data to eight.  Moreover, there are three gauge transformations on a 3-surface that specify the physically equivalence of every three solutions generated by different boundary data. These gauge symmetries leave five independent data. The last equivalence is due to the change of location of the boundary through the spacetime such that the same solution is generated\cite{wald2010general}. At the end, there remain four functions that can be freely specified on the boundary to generate physically distinct solutions of the equations of motion. These are the numbers of field and momenta degrees of freedom. Dividing by two, one concludes that there are two degrees of freedom at each point of the spacetime for the gravitational field. This matches with the degrees of freedom of a massless spin-2 field propagating in a flat spacetime. Although this function counting produces the expected result, there is no clue to which functions are the free data and which are determined by the constrains and so on, without an explicit gauge fixing. In the case of null-boundaries, even this crude counting is problematic and needs further investigation. We will come back to this matter in section \ref{nul case}.

	\subsection{Corner terms}
	To see how corner terms come out, let's consider two boundary segments $\mathcal B$ and $\mathcal B'$ which intersect on the corner $\mathcal C = \mathcal B \cap \mathcal B'$. Near this corner, we adopt a double foliation by ($\phi,\phi'$) such that $\phi =$ constant on $\mathcal B$ and $\phi' =$ constant on $\mathcal B'$. In addition, $n_a = N\,\nabla_a\phi$ and $n'_a = N'\,\nabla_a\phi'$ are unit normals to $\mathcal B$ and $\mathcal B'$, respectively. The advantage of using this adapted double foliation is that beside locating the boundary segments, it will give the foliation of each into sets of co-dimension two surfaces, with the corner $\mathcal C$ belonging to both sets. We consider the foliation of $\mathcal B$ with $\psi$ to be induced by the foliation $\phi'$ of the space time. Then, $m_a$, the unit normal to the co-dimension two surfaces in $\mathcal B$, can be written as
	\begin{align}
	m_a = \frac{n'_a - \epsilon (n\cdot n')n_a}{\sqrt{|\epsilon'-\epsilon(n\cdot n')^2|}}
	\end{align}
	in which $\epsilon=n\cdot n=\pm 1$, and $\epsilon'=n'\cdot n'=\pm 1$ are the normalization of unit normal vectors indicating the character of the boundaries $\mathcal B$ and $\mathcal B'$, respectively. With this realization of $m_a$, for the integrand of last integral in \eqref{BTnonnull}, we find
	\begin{align}
	m^a\,\deltabar u_a = m^a\,n^b\,\delta g_{ab} = \frac{n'^a - \epsilon (n\cdot n')n^a}{\sqrt{|\epsilon'-\epsilon(n\cdot n')^2|}}\,n^b\,\delta g_{ab} = \frac{n'^a\,n^b\,\delta g_{ab} -2\,n'^a\,\delta n_a}{\sqrt{|\epsilon'-\epsilon(n\cdot n')^2|}}
	\end{align}
	
	
	Similarly, the foliation of spacetime by $\phi$ induces a foliation in $\mathcal B'$, with unit normals to the co-dimension two surfaces
	\begin{align}
	m'_a = \frac{n_a - \epsilon' (n\cdot n')n_a}{\sqrt{|\epsilon'-\epsilon(n\cdot n')^2|}}
	\end{align}
	and the contribution to the corner term is
	\begin{align}
	m'^a\,\deltabar u'_a = m'^a\,n'^b\,\delta g_{ab} = \frac{n^a - \epsilon' (n\cdot n')n'^a}{\sqrt{|\epsilon'-\epsilon(n\cdot n')^2|}}\,n'^b\,\delta g_{ab} = \frac{n^a\,n'^b\,\delta g_{ab} -2\,n^a\,\delta n'_a}{\sqrt{|\epsilon'-\epsilon(n\cdot n')^2|}}
	\end{align}
	
	Finally, the addition of these two contributions gives
	\begin{align}
	m^a\,\deltabar u_a +m'^a\,\deltabar u'_a = \frac{2\,(n'^a\,n^b\,\delta g_{ab} -\,n'^a\,\delta n_a - \,n^a\,\delta n'_a)}{\sqrt{|\epsilon'-\epsilon(n\cdot n')^2|}} = \frac{-2\,\delta(n\cdot n')}{\sqrt{|\epsilon'-\epsilon(n\cdot n')^2|}} = \delta\vartheta\,
	\end{align}
	where the expression of $\vartheta$, for different cases of neighboring boundary segments, depends on their character and the value of $n\cdot n'$ as:
	\begin{align}
	\vartheta = \left\{\begin{array}{ll}
	-2\sinh^{-1}(n\cdot n') & \epsilon\,\epsilon' = -1 \\
	2\,\cos^{-1}(n\cdot n') & \epsilon\,\epsilon' = 1 \ \&\  |n\cdot n'|<1\\
	2\,\cosh^{-1}(n\cdot n') & \epsilon\,\epsilon' = 1 \ \&\ |n\cdot n'|>1
	\end{array}\right.
	\end{align}
	These results show that when the two neighboring segments of the boundary are of the same character, the corner term will be twice the parameter of the Lorentz transformation that converts one's normal to the other's on their joints (e.g the angle or the rapidity, depending on the value of $n\cdot n'$). In the case of two segments with different character, the result has found different interpretations as the boost parameter of the transformation $(n,m)\to(m',n')$ in \cite{Lehner2016} or giving an imaginary contribution to the action \cite{Neiman:2012fx}. 
	
	With this result, third term in \eqref{BTnonnull} of the two neighboring boundary segments contributes to make a total variation and a canonical pair on the corner between them:
	\begin{align}
	\int_{\mathcal C}\!\!\! \ud^{d-2} x \sqrt{|q|}\, (m^a\,\deltabar u_a +  m'^a\,\deltabar u'_a) =& \delta\left(\int_{\mathcal C}\!\!\! \ud^{d-2} x \sqrt{|q|}\,\vartheta\right) \nonumber\\
	&+\tfrac12\, \int_{\mathcal C}\!\!\! \ud^{d-2} x \,\sqrt{|q|}\,\vartheta\,q^{ab}\,\delta q_{ab}
	\end{align}
	Reading the boundary and corner action from total variation terms, one can write the proper action which has well-posed variational principle, in the aforementioned form of \eqref{action_general_form}. We have seen how a double-foliation of spacetime is essential to find the corner terms. We have also mentioned before that double-foliation can help to resolve the problem of restrictions on metric variations that keep the character of the boundary unchanged. In the next section, where we treat the null boundaries, we see the unavoidable need to a double-foliation.
	
	Before ending this section, let us summarize the method that we have used here. We have derived the boundary action and corner terms beside the canonical structure of Einstein gravity, with a systematic approach. The summary of the procedure is as follows\footnote{The Mathematica notebook of calculations based on this procedure is provided in the arxiv source files.}:
	
	\begin{enumerate}
		\item Varying the action to find the equations of motion and surface integral in terms of covariant derivative of metric variation $ \nabla_a\delta g_{bc} $:  \cref{boundaryInt}
		\item Decomposing  metric variation with respect to the boundary (similar to scalar-vector-tensor decomposition in metric perturbation theory): \cref{deltagdecompose}
		\item Taking derivative of the decomposed metric and decomposing the result.  (Here we recursively use a substitution $ n^a \nabla_c T_{\cdot\cdot a\cdot\cdot}=-T_{\cdot\cdot a\cdot\cdot} \nabla_c n^a $, for any tangential tensor $ T $ and use the expression \eqref{delndecompose} for  $  \nabla_c n^a $): \cref{deldeltagdecompose2}.
		\item Calculating variation of geometric object associated with the boundary similar to last step, as in \eqref{deltaK}, and solving  $ \nabla_a\delta g_{bc} $  in terms of them, which ends in final expression: \cref{finalCDG}.
		\item Substituting the final result of the last step in the result of first step and disintegrating the surface term to total variations and canonical pairs.  
	\end{enumerate}  
	In the next section, we follow the same procedure in treating the surface term of the Einstein theory in the case that the boundary is of null character.
\section{Null Boundary Segments}\label{nul case}
Suppose that a segment $\mathcal B$ of the spacetime boundary is a null hypersurface characterized by $\phi_0=0 $. Then, $\nabla_a\phi_0 \nabla^a\phi_0 = 0$ indicates that the normal vector to the null surface is also tangent to it. As a consequence, the induced metric becomes degenerate and constructing a projector to the null surface just from its normal is not possible; particularly, if someone \textit{naïvely} wants to use a similar prescription like the non-null case and take $\Pi^a{}_b = \delta^a{}_b + \alpha\,\ell^a\,\ell_b$ as projector, it reveals that for any $v\notin \mathcal T_p(\mathcal B)$, $\ell_a(\Pi^a{}_b\,v^b)=\ell_a\,v^a\ne 0$ which means that $\Pi^a{}_b\,v^b \notin \mathcal T_p(\mathcal B)$ and $\Pi^a{}_b$ is not a projector onto the null hypersurface. The standard remedy is to introduce an auxiliary vector $k^a$ which lays out of the hypersurface and therefor $\ell_a \, k^a \ne 0$ (for more details we refer the reader to \cite{Gourgoulhon2006,Jezierski:2003hh}). 

\subsection{Normal frame and double foliation}
We choose $\ell_a$ to be normal to the null boundary. Therefore, on the null boundary $ \ell_a=A\nabla_a\phi_0 $. We introduce the auxiliary null form $k_a$ and take the normalization of these null forms to be everywhere
\begin{align}
\ell_a\,\ell^a =0\ ,\ k_a\,k^a = 0 \quad\text{and}\quad \ell_a\,k^a=-1\,. \label{lknormalization}
\end{align}
Therefore, $\nabla_a(\ell^2) = \nabla_a(k^2) = \nabla_a{(\ell\cdot k)}=0$. 

With the aid of $\ell_a$ and $k_a$, we define the projector (see \cite{Gourgoulhon2006})
 \begin{align}
q^a{}_b = \delta^a{}_b + \ell^a\,k_b + k^a\,\ell_b
\end{align}
that essentially projects spacetime vectors onto the co-dimension two surface $\mathcal S$, to which $\ell_a$ and $k_a$ are orthogonal. 

A systematic approach to define these co-dimension two surfaces is to use a double-foliation by two scalar fields $ (\phi_0,\phi_1) $. The intersection of level surfaces of $\phi_0$ and $\phi_1$ are the co-dimension two surfaces $\mathcal S$. In this foliation $ \ell_a $ and $ k_a $ can be expanded generally as :  
\begin{align}
&\ell_a=A \,\nabla_a\phi_0+B\, \nabla_a\phi_1 \label{lfoliation}\\
&k_a=C\, \nabla_a\phi_0+D\, \nabla_a\phi_1 \label{kfoliation}
\end{align}
Three of four coefficients in the above expansions are determined by normalization conditions \eqref{lknormalization} and one remains free, due to the rescaling gauge freedom $(\ell_a \to \alpha\ell_a,k_a\to\frac1\alpha\kappa_a)$. On the boundary, we set the coefficient $ B=0 $, so that the boundary is a level surface of $\phi_0$ and we have $\ell_a  \overset{\mathcal B}=A\,\nabla_a\phi_0  $ . 

Similar to the non-null case, we decompose $\nabla_a \ell_b$ and $\nabla_a k_b$ as follows:
\begin{align}
\nabla_a \ell_b = \Theta_{ab} + \omega_a\,\ell_b + \ell_a\,\eta_{b} - k_a\,a_b - \kappa\,k_a\,\ell_b - \bar{\kappa}\,\ell_a\,\ell_b \label{delldecomposition}\\
\nabla_b k_b = \Xi_{ab} - \omega_a\,k_b + k_a\,\bar{\eta}_b - \ell_a\,\bar a_b + \kappa\,k_a\,k_b + \bar{\kappa}\,\ell_a\,k_b \label{delkdecomposition}
\end{align}
with definitions
\begin{equation}\label{BGOdef}
\begin{aligned}
&\Theta_{ab} = q^c{}_a\,q^d{}_b\,\nabla_a\ell_b \ &&, \quad \Xi_{ab} = q^c{}_a\,q^d{}_b\,\nabla_a\ k_b \\
&\eta_a = - q^c{}_a\,k^b\,\nabla_b\ell_c \ &&, \quad \bar{\eta}_b = - q^c{}_a\,\ell^b\,\nabla_b k_c \\
&\omega_a = - q^c{}_a\,k^b\,\nabla_c\ell_b = q^c{}_a\,\ell^b\,\nabla_c k_b \\
&a_a = q^c{}_a\,\ell^b\,\nabla_b\ell_c \ &&, \quad \bar a_a = q^c{}_a\,k^b\,\nabla_b k_c \\
&\kappa = -\ell^a\,k^b\,\nabla_a\ell_b =  \ell^a\,\ell^b\,\nabla_a k_b \ &&, \quad \bar\kappa = k^a\,\ell^b\,\nabla_a k_b = -k^a\,k^b\,\nabla_a\ell_b
\end{aligned}
\end{equation}
where $\Theta_{ab}$ and $\Xi_{ab}$ are extrinsic curvatures of $\mathcal S$, $\omega_a$, $\eta_a$ and $\bar{\eta}_a$ are twists, $a_a$ and $\bar a_a$ are tangent accelerations of $\ell^a$ and $k^a$ to $\mathcal S$, respectively, and $\kappa$ and $\bar\kappa$ are in-affinity parameters\footnote{The quantity $\kappa$ is called surface gravity in the case of a black hole horizon null surface.}. Note that we have  used the defining conditions $\nabla_a(\ell^2) = \nabla_a(k^2) = \nabla_a{(\ell\cdot k)}=0$ in decompositions \eqref{delldecomposition} and \eqref{delkdecomposition}.

\subsection{Variations and their decompositions}
Similar to the non-null case, we decompose the variation of metric into a tensor, two vectors and three scalars defined on the co-dimension two surface $ \mathcal S $ as follows:
\begin{align}
&\deltabar q_{ab} = q^c{}_a\,q^d{}_b\,\delta g_{cd} \nonumber\\
&\deltabar u_{1a} =- q^b{}_a\,\ell^c\,\delta g_{bc} \quad,\quad
\deltabar u_{2a} =- q^b{}_a\,k^c\,\delta g_{bc}\quad\nonumber\\
&\deltabar \mu_{1} =\ell^a\,\ell^b\,\delta g_{ab} \qquad\ \,,
\quad\deltabar \mu_{2} =k^a\,k^b\,\delta g_{ab} \quad,\quad
\deltabar \mu_{3} =\ell^a\,k^b\,\delta g_{ab}\quad
\end{align}
So, the variation of metric is expanded as:
\begin{align}
\delta g_{ab} &= \deltabar q_{ab} + 2 k_{(a} \deltabar u_{1b)} + 2\ell_{(a} \deltabar u_{2b)} + k_a k_b \deltabar \mu_1 + \ell_a\ell_b \deltabar \mu_2 +2 \ell_{(a} k_{b)}\deltabar \mu_3 \,.\label{deltag}
\end{align}
Again, we have to remind that in the right hand side of the above expressions
$ \deltabar $  does not introduce a variation of some function especially: $ \deltabar q_{ab}\neq\delta(q_{ab}) $.

It is mandatory that the variation keeps the normalization conditions of the frame forms $ \ell $ and $ k $ unchanged. Thus, from normalization condition of $\ell$, we have:
\begin{align}
0=\delta (\ell_a\,\ell^a) = 2 \ell^a\delta \ell_a + \ell_a \ell_b \delta(g^{ab}) = 2 \ell^a\delta \ell_a - \deltabar \mu_1 \,.\label{eq1}
\end{align}
Similarly, from normalization condition of $k$:
\begin{align}
0=\delta (k_a\,k^a) = 2\,k^a\,\delta k_a - k^a\,k^b\,\delta g_{ab}= 2\,k^a\,\delta k_a - \deltabar \mu_2 \,.
\end{align}
Finally from the choice of dot product between $\ell$ and $k$:
\begin{align}
0=\delta (\ell_a\,k^a) = \ell^a\,\delta k_a + k^a\,\delta \ell_a -\ell^a\,k^b\,\delta g_{ab}= \ell^a\,\delta k_a + k^a\,\delta \ell_a - \deltabar \mu_3 \,.
\end{align}
Using the facts that $q^a{}_b\,\delta\ell_a=0$ and $q^a{}_b\,\delta k_a=0$ \footnote{See \cref{deltal2,deltak2} of Appendix \ref{appendix:variation of geometric objects}}, the general solution of the above three equations are:
\begin{align}
\delta \ell_a &= -(\deltabar \mu_3 + \deltabar\beta) \ell_a  -\frac12\,\deltabar \mu_1\,k_a \label{deltal}\\
\delta k_a &= \deltabar\beta \,k_a -\frac12\,\deltabar \mu_2\,\ell_a \label{deltak}
\end{align}
for arbitrary function $\deltabar\beta$, which is related to the rescaling gauge freedom. In appendix \ref{appendix:variation of geometric objects}, we derive the above equations in a double foliation set-up and find explicitly the relations between metric variations and foliation coefficients. There, we find that $\deltabar\beta= \delta  \ln\! D$. 

Although $\ell_a$ and $k_a$ remain null under variations, the boundary hypersurface in general dose not remain null. To see this, firstly note that $\deltabar \mu_1 \propto \delta B$ (as it is shown in appendix \ref{appendix:variation of geometric objects}). This fact, alongside the expansion \eqref{lfoliation}, indicates that $\ell_a$ will no longer remain normal to the hypersurface, if $\deltabar \mu_1 \ne 0$. Secondly, we note that the character of the boundary is defined by the norm of $\nabla_a \phi_0$. So, under a general variation of the metric it changes like:
\begin{align}
\delta(\nabla^a \phi_0 \nabla_a \phi_0) = -\delta g^{ab} \nabla_a \phi_0 \nabla_b \phi_0 = -\frac1{A^2}\,\delta g^{ab}\,\ell_a\ell_b = - \frac{\deltabar \mu_1}{A^2} \,.
\end{align}
This shows that only if we take $\deltabar \mu_1=0$, the variations will be restricted to keep the boundary null. However, we don't choose this restriction and allow the boundary changes its character. 

\subsection{Disintegration of null surface term}
Now we compute the surface term \eqref{boundaryInt} on a null hypersurface. We can, as we have done in the previous section for non-null boundaries, calculate $\nabla_a\delta g_{bc}$ and express it in terms of $\delta \Theta_{ab}, \delta \Xi_{ab},$ etc. This is what we will do in appendix \ref{Explicit}. But to obtain the boundary term in Einstein theory we can write directly the boundary term as
\begin{align}
2\,P^{abcd}\,\ell_a\,\nabla_d \delta g_{bc} =& \ell^a\,\nabla^b \delta g_{ab} - \ell^a\,\nabla_a\delta g^b{}_b\nonumber\\
=&-\ell^a\nabla_a \deltabar q^b{}_b - \nabla_a \deltabar u_1^a - k^a\,\nabla_a\deltabar \mu_1 +\ell^a\,\nabla_a\deltabar \mu_3 \nonumber\\
&+ \left(-\deltabar q_{ab}-2\,k_{(a}\,\deltabar u_{1b)} - \,\ell_{a} \, \deltabar u_{2b}  -k_a\,k_b\,\deltabar \mu_1 - (\ell_a\,k_b + g_{ab})\deltabar \mu_3 \right)\nabla^a\ell^b \nonumber\\
&- \deltabar \mu_1\,\nabla_a k^a \label{Pdeldeltag1}
\end{align}
From eq\eqref{deltaTheta} in appendix \ref{appendix:variation of geometric objects} we find
\begin{align}
\delta \Theta = q^{ab}\,\delta \Theta_{ab} - \Theta^{ab}\,\deltabar q_{ab} =& \tfrac12\,\ell^a\nabla_a(\deltabar q^b{}_b) + \nabla^a\deltabar u_{1a}\nonumber\\
& - (\omega^a - \eta^a)\,\deltabar u_{1a} + a^a\,\deltabar u_{2a}	- \tfrac12\,\Xi\,\deltabar \mu_1 - \Theta\,\delta  \ln\! D\,, \label{tracedeltaTheta}
\end{align}
also from \eqref{deltakappa} for $\delta\kappa$ we have:
\begin{align}
\delta \kappa =&  \tfrac{1}{2} \mathit{k}^{a} \nabla_{a}(\deltabar \mu_{1}) -  \ell^{a} \nabla_{a}(\deltabar \mu_{3}) -  \ell^{a} \nabla_{a}(\delta \ln\! D)\nonumber\\
&+(\omega^{a}-\eta^a)\deltabar u_{1a}  - \mathit{a}^{a} \deltabar u_{2a} -  \tfrac{1}{2} \overline{\kappa} \deltabar \mu_{1}  -   \kappa\,\delta  \ln\! D  \,.
\end{align}
Using the above equation for $\delta\Theta$ and $\delta\kappa$, the expression \eqref{Pdeldeltag1} reduces to
\begin{align}
2\,P^{abcd}\,\ell_a\,\nabla_d \delta g_{bc} =&  -2\,\delta (\Theta + \kappa) \nonumber\\
& - \Theta_{ab}\, \delta q^{ab} + (\omega^a+\eta^a)\deltabar u_{1a} - \mathit{a}^{a} \deltabar u_{2a} \nonumber\\
&+ \overline{\kappa}\, \deltabar \mu_{1} -   \Theta\,\deltabar \mu_{3} - 2  (\Theta +\kappa) \,\delta \ln\! D \nonumber\\
&+ \nabla_{a}(\deltabar u_{1}{}^{a})-  \ell^{a} \nabla_{a}(\deltabar \mu_{3}) - 2\, \ell^{a} \nabla_{a}(\delta  \ln\! D)\,,
\end{align}
where we have used the fact that $q^c{}_a\,q^d{}_b\,\deltabar q_{cd} = \delta q_{ab} $. 
For the term $ \nabla_a \deltabar u^{1a} $ we have:
\begin{align}
\nabla_a \deltabar u^{1a} = (q^{ab} + \ell^a\,k^b + k^a\ell^b)\nabla_a \deltabar u_{1b} = -(\eta^a+\bar{\eta}^a)\deltabar u_{1a} + \mathcal D^a(\deltabar u_{1a})
\end{align}
where we have used $\ell^b\,\nabla_a (\deltabar u_{1b})= - \deltabar u_1{}^b\,\nabla_a \ell_b$ and $k^b\,\nabla_a (\deltabar u_{1b})= - \deltabar u_1{}^b\,\nabla_a k_b$. Using this we obtain: 
\begin{align}
2\,P^{abcd}\,\ell_a\,\nabla_d \delta g_{bc} =&  -2\,\delta (\Theta + \kappa) \nonumber\\
& - \Theta_{ab}\, \delta q^{ab} + (\omega^a-\bar\eta^a)\deltabar u_{1a} - \mathit{a}^{a} \deltabar u_{2a} \nonumber\\
&+ \overline{\kappa}\, \deltabar \mu_{1} -   \Theta\,\deltabar \mu_{3} - 2  (\Theta +\kappa) \,\delta \ln\! D \nonumber\\
&+ \mathcal D^a(\deltabar u_{1a})-  \ell^{a} \nabla_{a}(\deltabar \mu_{3}) - 2\, \ell^{a} \nabla_{a}(\delta \ln\! D)\,.\label{nullBT1}
\end{align}

Thus the surface term \eqref{boundaryInt} yields to be
\begin{align}
\Theta_{\mathcal B}=&\int_{\mathcal B}\!\!\! \ud^{d-1} x \,\frac{\sqrt{-g}}A \big\{ -2\,\delta (\Theta + \kappa) \nonumber\\
& - \Theta_{ab}\, \delta q^{ab} + (\omega^a-\bar\eta^a)\deltabar u_{1a} - \mathit{a}^{a} \deltabar u_{2a} \nonumber\\
&+ \overline{\kappa}\, \deltabar \mu_{1} -   \Theta\,\deltabar \mu_{3} - 2  (\Theta +\kappa) \,\delta \ln\! D \nonumber\\
&+ \mathcal D^a(\deltabar u_{1a})-  \ell^{a} \nabla_{a}(\deltabar \mu_{3}) - 2\, \ell^{a} \nabla_{a}(\delta  \ln\! D)\big\}\,.
\end{align}
In the double-foliation ($ \phi_0,\phi_1 $), on the null boundary we have  $\sqrt{-g}= AD\sqrt q$ and $a_a = 0$ (as is shown in appendix \ref{GDF}). Using these relations, we can rewrite the above result as
\begin{align}
\Theta_{\mathcal B}=& -2\,\delta\left(\int_{\mathcal B}\!\!\! \ud^{d-1} x \, D\,\sqrt q (\Theta + \kappa)\right) \nonumber\\
&- \int_{\mathcal B}\!\!\! \ud^{d-1} x \,D\,  \sqrt q\big\{(\Theta^{ab} -q^{ab}\,(\Theta + \kappa))\delta q_{ab}  + (\bar\eta^a - \omega^a)\deltabar u_{1a}  - \overline{\kappa}\, \deltabar \mu_{1} \nonumber\\
& \qquad\qquad\qquad\quad - \deltabar u_{1a}\,\mathcal D^a(\ln D) + 2\, \ell^{a} \nabla_{a}(\delta  \ln\! D)\big\}\nonumber\\
&+\int_{\partial\mathcal B}\!\!\! \ud^{d-2} x \,D\,  \sqrt q\,s^a(\deltabar \mu_3\,\ell_a + \deltabar u_{1a})\label{ThetaBnull}
\end{align}
where we have used integration by parts 
\begin{align}
&\int_{\mathcal B}\!\!\! \ud^{d-1} x \,D\,  \sqrt q\, \ell^a\,\nabla_a(\deltabar \mu_3) = - \int_{\mathcal B}\!\!\! \ud^{d-1} x \,D\,  \sqrt q\,\deltabar \mu_3\,\Theta^a_a + \int_{\partial\mathcal B}\!\!\! \ud^{d-2} x \,D\,  \sqrt q\,s^a\,\ell_a\,\deltabar \mu_3
\\
&\int_{\mathcal B}\!\!\! \ud^{d-1} x \,D\,  \sqrt q\, \mathcal D^a (\deltabar u_{1a}) = - \int_{\mathcal B}\!\!\! \ud^{d-1} x \,D\,  \sqrt q\,\deltabar u_{1a}\,\mathcal D^a(\ln D) + \int_{\partial\mathcal B}\!\!\! \ud^{d-2} x \,D\,  \sqrt q\,s^a\,\deltabar u_{1a}
\end{align}

If the boundary segment $\mathcal B$ has the topology $S^{d-2}\times \mathbb R$, where $S^{d-2}$ is a compact manifold, then $s_a = \tfrac1D k_a$ and we will have for the last line in \eqref{ThetaBnull} 
\begin{align}
&\int_{\partial\mathcal B}\!\!\! \ud^{d-2} x \,  \sqrt q\,k^a(\deltabar \mu_3\,\ell_a + \deltabar u_{1a}) = - \int_{\partial\mathcal B}\!\!\! \ud^{d-2} x \,  \sqrt q \,\deltabar \mu_3\nonumber\\& = - \delta( \int_{\partial\mathcal B}\!\!\! \ud^{d-2} x \,  \sqrt q \,\ln \sqrt H) +\tfrac12  \int_{\partial\mathcal B}\!\!\! \ud^{d-2} x \sqrt{q}(\sqrt H q^{ab})\delta q_{ab}
\end{align}
where $H $ is the determinant of transverse metric and the relation of its variation to $ \deltabar \mu_3 $ has been proven in appendix \ref{appendix:variation of geometric objects}.

If we fix the scaling gauge by choosing $ D=1 $\footnote{In section \ref{general case} we present an expression without fixing the scaling gauge.}, we get the final expression for the surface term:
\begin{equation}
\begin{aligned}
\Theta_{\mathcal B}=& -2\,\delta\left(\int_{\mathcal B}\!\!\! \ud^{d-1} x \, \,\sqrt q (\Theta + \kappa)\right) - \delta\left( \int_{\partial\mathcal B}\!\!\! \ud^{d-2} x \,  \sqrt q \,\ln \sqrt H\right) \\
&- \int_{\mathcal B}\!\!\! \ud^{d-1} x \,  \sqrt q\left[\left(\Theta^{ab} -q^{ab}\,(\Theta + \kappa)\right)\delta q_{ab}   - 2\,\omega^a\,\deltabar u_{1a}  - \overline{\kappa}\, \deltabar \mu_{1}\right] \\
&+\tfrac12  \int_{\partial\mathcal B}\!\!\! \ud^{d-2} x\sqrt q (\sqrt H q^{ab})\delta q_{ab} \,.
\label{ThetaBnullfinal}
\end{aligned}
\end{equation}
The first line of the above expression hints to which boundary action is needed in order to have a well-posed variational principle. The last two lines give the canonical structure on null boundary.

In appendix \ref{appendix:variation of geometric objects}, we will show $ \deltabar u_{1a} = -\delta \beta_{1a} $ and $\deltabar \mu_1 = -2\,\delta B$. These mean that $ \deltabar u_{1a} $ and $ \deltabar \mu_1 $ are indeed variation of the functions defined in the double-foliation of spacetime. Substituting these relations in the second line of \eqref{ThetaBnullfinal}, we obtain the final expression for the canonical structure
\begin{align}
\Omega = -\int_{\mathcal B}\!\!\! \ud^{d-1} x \,  \sqrt q\left[\left(\Theta^{ab} -q^{ab}\,(\Theta + \kappa)\right)\delta q_{ab}   + 2\,\omega^a\,\delta \beta_{1a}  + 2\, \overline{\kappa}\, \delta B \right]\,.
\end{align}
The null canonical structure derived here coincides with previous studies, except for an important extra term $\bar{\kappa}\,\delta B$ that emerges here. This term contains a new canonical pair  of the theory: $ (B\,,\, 2\,\bar{\kappa}) $. This emergence is due to the removal of the restriction on variations, so that the boundary can changes its character. The new configuration variable  $B$ is one of the coefficients in equation \eqref{lfoliation}, and its nonzero value makes the boundary $\mathcal B$ to become non-null. The conjugate momentum of this variable reads to be $\bar{\kappa}$, the normal acceleration of transverse vector $k^a$.

With the complete null canonical structure in hand, we are at the stage of evaluating the number of degrees of freedom of the gravitational field in spacetimes with null-boundaries. Canonical pairs are $(q_{ab}\,,\,\Theta_{ab}-q_{ab}(\Theta + \kappa))$ , $ (\beta_{1a}\,,\, 2\,\omega_a) $ and $ (B\,,\, 2\,\bar{\kappa}) $, which sum up to $ 2\times (3 + 2 +1) = 12 $ functions. This number matches with that of a freely specifiable function in non-null case, before considering the constraint equations and gauge redundancies. We note that the new canonical pair $ (B\,,\, 2\,\bar{\kappa}) $ that is found here, is necessary for this matching. To reduce the number of these freely specifiable functions on the null boundary, we need to count the constraint equations and gauge redundancies imposed on them. The equation of motions decomposed with respect to the null boundary are derived in Appendix \ref{app:decompositionEOM}. Separating these equations into dynamical and constraint equations, in this case, needs more careful treatment. We will address this problem in a future work \cite{Jafari2018b}.

\section{Revisiting the Non-Null: General Boundaries}\label{general case}
As we have seen in the previous section, some variations change the character of boundary from being null to time-like or space-like. Naturally, one may expect the reverse to be also true, i.e. the space-like or time-like boundaries to become null under variation. We have seen that the standard set-up used in section \ref{non-null case} fails to include these kinds of situations. In this section, we benefit from the double foliation of spacetime, introduced in the previous section, to be able to consider such variations. But before that, we briefly mention another choice by using the non-normalized $v_a = \nabla_a \phi$ as the normal of the boundary for decomposition of variations. With this choice, the projector $h_{ab}$ becomes
\begin{align}
h_{ab}  = g_{ab} - \tfrac1{\mathcal N} \,v_a\,v_b \label{projector}
\end{align}
where $\mathcal N = v^a v_a$.
Similar calculations  to those presented in section \ref{non-null case} leads to the following integrand for the surface  term:
\begin{align}
2\,\sqrt{-g}\,P^{abcd}\,v_a\,\nabla_d (\delta g_{bc}) = & \sqrt{-g}\left( 2\,h^{ab}\,\delta \mathcal K_{ab} - \mathcal K^{ab}\,\deltabar h_{ab} - \mathcal N\,\mathcal K \,\deltabar \mu - \,D^a(\mathcal N\,\deltabar u_a)
\right)\nonumber\\
=& \, 2\,\delta\left(\sqrt{-g}\, \mathcal K\right) +\sqrt{-g}\, (\mathcal K^{ab}- \mathcal K \,h^{ab})\,\deltabar h_{ab} - \sqrt{-g}\,\,D^a(\mathcal N \deltabar u_a)\,, \label{surfaceterm2}
\end{align}
where here $ \mathcal{K}_{ab}=h_a{}^{c}h_{b}{}^{d}\,\nabla_cv_d $ and
$\deltabar\mu = v^a\,v^b\,\delta g_{ab} = -\delta\mathcal N/\mathcal N^2 = \delta(\frac1{\mathcal N})$. 

Note that here under variation we have:
\begin{equation}
(v^av_a)'=(v^av_a)+\delta(v^av_a)=\mathcal{N}+\delta\mathcal{N}
\end{equation}
so  that by $\delta \mathcal{N}=-\mathcal{N} $, the variations can turn the character of the boundary into null. Although the restriction on variations is removed in this approach and the total variation term is derived\footnote{Note the difference with the standard $ \sqrt{h}K $ term.}, but realizing the canonical structure is problematic. This is because $\delta h_{ab} = h'_{ab} - h_{ab}$ is not meaningful when the boundary becomes null. Therefore $\deltabar h_{ab}$ in the expression \eqref{surfaceterm2}, in general, can not be replaced by a true variation of the induced metric. 

Now we return to applying the double-foliation approach and see how the aforementioned problems are resolved there. In double-foliation of spacetime that we use here, we take the boundary to be a $\phi_0=cte$ surface, regardless of its character which can be space-like, time-like or null. We choose the frame to be the same as one in the previous section with $\ell_a$ and $k_a$ having the normalization conditions \eqref{lknormalization} and expansion relations \eqref{lfoliation} and \eqref{kfoliation}.

The procedure here is very similar to what we have done in the previous section, apart from the normal that we find from \cref{lfoliation,kfoliation}.  to be
\begin{equation}\label{generalnormal}
v_a=\nabla_a\phi_0=\tfrac{1}{\sqrt{H}} (D\ell_a-Bk_a)\,.
\end{equation}
Unlike the previous section that the boundary was supposed to be null and therefore $B=0$ on the boundary, here in general depending on the values of $B$ and $D$, the boundary can be of any character. With a similar calculation we find\footnote{We haven't repeated the analysis of the corner terms, which is similar to that in section \ref{non-null case}}.:
\begin{align}
2\,\sqrt{-g}\,P^{abcd}\,v_a\,\nabla_d \delta g_{bc} =&\, -\sqrt{q}\,\big(2\, D\, \delta \Theta  + 2\, D\, \delta \kappa + 2\, \delta D\, \kappa   - 2\, ‌\,B \delta \Xi + 2\, ‌B\, \delta \bar{\kappa} + 2\, \bar{\kappa} \,\delta ‌B\nonumber\\
 &+  (D \, \Theta^{ab}  - ‌B \,\Xi^{ab})\delta q_{ab} - 2\, \omega^{a}\,\delta \beta_{1}{}_{a}\big)\label{GB1}\\
 =&- \,2\,\delta\left(\sqrt q (D(\Theta + \kappa)-B(\Xi-\bar{\kappa}))\right) \nonumber\\
 &-\sqrt q \big[ D(\Theta^{ab} - q^{ab}(\Theta +\kappa) ) - B (\Xi^{ab} - q^{ab} (\Xi - \bar{\kappa}))\big]\delta q_{ab} \nonumber\\
 &+ 2\, \omega^{a}\,\delta \beta_{1}{}_{a} + 2\,\Theta\,\delta D  - 2\,\Xi\,\delta B \label{GB2}
\end{align}
It can easily be seen in \eqref{GB1} that for $ B=0 $ (but not $ \delta B=0 $ ), we get the result of previous section consistently. In the alternative form of \eqref{GB2}, that total variations are produced, in order to rederive the null result, one should note that, variations are defined on the bulk and the boundary-defining relation $ B=0 $ has to be applied after taking the variation. Therefore, for example we have: $\delta(B\,\bar{\kappa}) = B\,\delta\bar{\kappa} + \bar{\kappa}\,\delta B \overset{\mathcal B}=\bar{\kappa}\,\delta B $.

From \eqref{GB2}, the proper boundary action reads to be
\begin{equation}
2\,\int_{\mathcal B}\!\!\! \ud^{d-1} x \, \,\sqrt q [D(\Theta + \kappa)-B(\Xi-\bar{\kappa})]  \label{CTG}
\end{equation}
in which the second term on the null boundary vanishes. But on the non-null boundaries, where $B\ne 0$, with the projector \eqref{projector} in which $\mathcal N = -\frac{2BD}H$, we can rewrite \eqref{CTG} as $2\int_{\mathcal B} \sqrt{-g}\,\mathcal K$, which coincides with the boundary action that \eqref{surfaceterm2} suggests. Therefore \eqref{CTG} provides a general boundary action for the Einstein-Hilbert theory, which is derived in a double-foliation formalism. 

\section{Discussion}
In this paper, we have organized a systematic and robust approach that can be used to study the variational principle and to derive the canonical structure of the gravitational theory whose dynamical field is the spacetime metric and its action consists of derivatives of metric. As a first example, we have used this approach for General Relativity when the boundaries of spacetime are non-null. We have reviewed and rederived the well-known results in this case, like the boundary and corner actions and the canonical structure. Then, we turned to the same theory on spacetimes whose boundaries contain null segments. We have found the recently derived results in this case with a different approach, besides a new canonical pair that was missing in previous works, due to a restriction on variations. This data can be important in problems where the boundary of the spacetime region under study changes its character. Examples are dynamical black hole horizons, apparent horizons and other dynamical trapping horizons in quasi-local descriptions of black holes (see for example \cite{hayward2013black , Andersson:2005gq}).

We have shown that double-foliation of spacetime provides a clear picture and powerful tool, not only when one deals with null boundary segments or corners, but also in deriving general results for boundaries of arbitrary character with unrestricted variations. We have expressed the details of general double-foliation of spacetime and variations in that framework in appendices \ref{GDF} and \ref{appendix:variation of geometric objects}. Derivative of variation of metric is derived explicitly, in this framework in appendix \ref{Explicit}. Finally, in appendix \ref{app:decompositionEOM}, we have decomposed completely the Riemann and Einstein tensors.

The canonical structure that is derived in covariant formulation, coincides with canonical pairs that are found in Hamiltonian formulation of Einstein theory on spacetimes with non-null boundaries. There remains the task of comparing the canonical structure in a null boundary case with the Hamiltonian approach. In order to provide a counting of degrees of freedom on null boundaries, one also needs the classification of various components of the equations of motion. The Hamiltonian formulation in double-null foliation of spacetime has been studied before \cite{Hayward1993,Vickers2011}. But a general double-foliation that fulfills the required framework for unified treatment of general hypersurfaces is missing. We will return to this problem in our next work.

For higher-derivate theories of gravity, such as Lanczos-Lovelock theory, there have been several attempts to find the boundary and corner actions (see  \cite{Chakraborty2017,Chakraborty:2018dvi,Cano:2018ckq}). A future study is to investigate this problem with the systematic approach that is developed in this paper. The canonical structure of these theories has not been derived in general, using the variational methods. Our approach may help to step forward in that direction too. A related problem is the study of conserved charges on general boundaries. The celebrated covariant phase space method that is proposed in \cite{Iyer:1994ys} and used for conserved charges at null infinity in \cite{Wald:1999wa}, was applied to finite bounded regions of spacetime in \cite{Hopfmuller:2018fni,Chandrasekaran:2018aop}. A relevant question would be the role of the new canonical pair found in this paper in the symplectic form and the resulting conserved charges.

\section*{Acknowledgement}
The authors would like to thank M. M. Sheikh-Jabbari, F. Hopfmüller, K. Parattu, S. Chakraborty for useful discussions and their helpful comments. G. J. and S. A. specially thank to M. Alishahiha for motivating this study. The abstract tensor calculation in this paper has been carried out by Mathematica package xAct \cite{MARTINGARCIA2008597,Nutma:2013zea}. 
\newpage
\appendix
\section{General Double Foliation}\label{GDF}
One of the key elements in our analysis is the usage of a double foliation of spacetime which is general in any sense except that it is adapted to the  boundary segment $\mathcal B$; i.e. the null boundary hypersurface is a leaf of one of the foliations. These foliations are set up by $ \phi_0 $ and $ \phi_1 $, which are scalar fields on the spacetime near the boundary. The null boundary segment $\mathcal B$ is a level surface of $\phi_0$. The intersection of any two level surfaces of $\phi_0$ and $\phi_1$ is a $d-2$ dimensional surface called $\mathcal S$.

We define two null normals to $\mathcal S$ as 
\begin{align}\label{lkdd}
&\ell_a=A\, \nabla_a\phi_0+B\, \nabla_a\phi_1\nonumber\\&
k_a=C\, \nabla_a\phi_0+D\, \nabla_a\phi_1 \,,
\end{align}
On the null boundary we set $B=0$, so that $ \ell_a \overset{\mathcal B}= A\,\nabla_a\phi_0 $. This implies that $ \ell_a$ is normal to the null boundary.

In a double foliation framework, we can write the spacetime metric as
\begin{align}
g_{ab}\,dx^a\,dx^b = H_{ij}\,d\phi^i\,d\phi^j + q_{AB}(d\sigma^A + \beta_i^A\,d\phi^i)(d\sigma^B + \beta_j^B\,d\phi^j)\,,
\end{align}
where $i,j\in\{0,1\}$ and $A,B\in\{2,\ldots ,d-1\}$, $\sigma^A$ are coordinates on co-dimension two surface $\mathcal S$ and $\beta^A_i$ are shift vectors. The normal metric $H_{ij}$ consists of laps functions as
\begin{align}
H_{ij} = -\left(
\begin{array}{cc}
2  A {C} & {B} {C}+{A} {D} \\
{B} {C}+{A} {D} & 2 {B} {D} \\
\end{array}
\right).
\end{align}
We can compute the vectors $\ell^a$ and $k^a$, which are
\begin{align}
\ell^a &= \frac1{AD-BC}\left[B (\partial_0 - \beta_0)^a - A (\partial_1 - \beta_1)^a\right] \,,\\
k^a &= \frac1{AD-BC}\left[-D (\partial_0 - \beta_0)^a +C (\partial_1 - \beta_1)^a\right] \,,
\end{align}
and thus,
\begin{align}
&\partial_0{}^a = -C \, \ell^a - D\, k^a + \beta_0^a \quad,\quad \partial_1{}^a = -D \, \ell^a - B\, k^a + \beta_1^a \,.
\end{align}

We can write $\Theta_{ab} = q_a^c\,q_b^d\,\nabla_c \ell_d$ and $\Xi_{ab} = q_a^c\,q_b^d\,\nabla_c k_d$ in terms of $\mathscr L_{\partial_i} q_{AB}$ as follows
\begin{align}
\Theta_{AB} &= \frac1{2\sqrt H}\,(B\,\mathscr L_{\partial_0} q_{AB} -A\,\mathscr L_{\partial_1} q_{AB} -2\,B\,\mathcal D_{(A} \beta_{0 B)}+2\,A\,\mathcal D_{(A} \beta_{1 B)})\\
\Xi_{AB} &= \frac1{2\sqrt H}\,(-D\,\mathscr L_{\partial_0} q_{AB} +C\,\mathscr L_{\partial_1} q_{AB} +2\,D\,\mathcal D_{(A} \beta_{0 B)}-2\,C\,\mathcal D_{(A} \beta_{1 B)})
\end{align}
These relations are similar to the well-known relation for extrinsic curvature in $3+1$ decomposition.

We can also find the relation between twist variables $\omega_a$, $\eta_a$ and $\bar{\eta}_a$ in terms of laps functions as:
\begin{align}
\omega_a - \eta_a &= \tfrac1{\sqrt H}(-D\,\mathcal D_a A + C\,\mathcal D_a B) \overset{\mathcal B}{=} - \mathcal D_a (\ln A)\\
\omega_a + \bar\eta_a &= \tfrac1{\sqrt H}(-B\,\mathcal D_a C + A\,\mathcal D_a D) \overset{\mathcal B}{=} - \mathcal D_a (\ln D)
\end{align}
and the acceleration
\begin{align}
a_a = \tfrac1{\sqrt H}(-B\,\mathcal D_a A + A\,\mathcal D_a B) \overset{\mathcal B}{=} 0\,.
\end{align}

\section{Variation of Geometric Objects}\label{appendix:variation of geometric objects}

The null 1-forms $\ell_a$ and $k_a$ are varied as
\begin{align}
&\delta\ell_a=\delta A \,\nabla_a\phi_0+\delta B\, \nabla_a\phi_1\nonumber\\&
\delta k_a=\delta C \,\nabla_a\phi_0+\delta D\, \nabla_a\phi_1
\end{align}
which can expressed in terms of $ \ell_a $ and $ k_a $ by inverting \eqref{lkdd}. 
\begin{align}
&\delta
\ell_a=\frac{1}{A D-B C}\left[ (D\, \delta A-C\, \delta B)\,\ell_a- (B\, \delta A-A\, \delta B)\,k_a\right]\nonumber\\
&\delta k_a=\frac{1}{A D-B C}\left[ (A\, \delta D-B\, \delta C)\,k_a- (C\, \delta D-D\, \delta C)\ell_a\,\right]\,.
\end{align}
These relations on the null hypersurface, where $B=0$, results in
\begin{align}
&\delta\ell_a \overset{\mathcal B}=\frac{1}{A D}\left[ (D\, \delta A-C\, \delta B)\ell_a+ A\, \delta B\,k_a\right]\label{deltal2} \\
&\delta k_a \overset{\mathcal B}=\frac{1}{A D}\left[ A\, \delta D\,k_a- (C\, \delta D-D\, \delta C)\ell_a\right] \label{deltak2}
\end{align}
comparing with \cref{deltal,deltak} we find:
\begin{align}
&\deltabar \mu_1 \overset{\mathcal B}= - 2\, \frac{\delta B}{D} \label{deltamu1}\\&
\deltabar \mu_2 \overset{\mathcal B}= \frac{2}{A D}(C \delta D-D \delta C) \label{deltamu2}\\
&\deltabar \mu_3 \overset{\mathcal B}= -\frac{\delta D}{D} -\frac{\delta A}A +\frac{C\,\delta B}{A\,D} \label{deltamu3}\\
& \delta\beta \overset{\mathcal B}= \frac{\delta D}{D}
\end{align}
we notice that the expression for $ \deltabar \mu_3 $ can be written as $ \deltabar \mu_3=\delta(\ln \sqrt H) $, where $ H=(AD-BC)^2 $ is the determinant of transverse metric. It means that $ \deltabar \mu_3 $ is in fact variation of some function.

We can find the relation between $\deltabar u_1$ and variation of the shift vectors $\beta^A_i$ as
\begin{align}
\deltabar u_{1a} = q_a{}^{b}\,\ell^c\,\delta g_{bc} = \tfrac1{\sqrt H}\,( B\,\delta \beta_{0a} - A\,\delta \beta_{1a}) \overset{\mathcal B}= -\frac{\delta \beta_{1a}}D
\end{align}

Using definitions in \eqref{BGOdef}, and \eqref{deltag}, we can find the following relation for $\delta\Theta$:
\begin{align}
\delta\Theta_{ab} =& \tfrac{1}{2} \Theta_{a}{}^c\,\deltabar q_{bc} 
+  \left(\mathit{k}_{a} \Theta_{b}{}^c- \tfrac{1}{2}  (\omega_{a}+ \overline{\eta}_{a})\delta_b^c\right)\deltabar u_{1c}  +  (\ell_{a} \Theta_{b}{}^c+\tfrac{1}{2} \mathit{a}_{a}\delta_b^c)\deltabar u_{2c} \nonumber\\
&+ \tfrac{1}{4}\, \Xi_{ab}\, \deltabar  \mu_{1}  -  \tfrac12\Theta_{ab}\, \delta \ln\! D   + \tfrac14\,q^c{}_{a}\, q^d{}_{b}\, \ell^{e}\, \nabla_{e}\deltabar q_{cd} + \tfrac12\,\mathcal D_a(\deltabar u_{1b}) + (a\leftrightarrow b)\,, \label{deltaTheta}
\end{align}
where we also used some relations in \eqref{BPrelations} of next appendix. Similarly for $\delta\Xi$ we get:
\begin{align}
\delta\Xi_{ab} =& \tfrac{1}{2} \Xi_{a}{}^c\,\deltabar q_{bc} +  (k_{a} \Xi_{b}{}^c+\tfrac{1}{2} \mathit{a}_{a}\delta_b^c)\deltabar u_{1c}
+  \left(\mathit{\ell}_{a} \Xi_{b}{}^c+ \tfrac{1}{2}  (\omega_{a}- \eta_{a})\delta_b^c\right)\deltabar u_{2c} \nonumber\\
&  + \tfrac{1}{4}\, \Theta_{ab}\, \deltabar \mu_{2}  -  \tfrac12\Xi_{ab}\, (\deltabar \mu	_3+\delta \ln\! D)   + \tfrac14\,q^c{}_{a}\, q^d{}_{b}\,k^{e}\, \nabla_{e}\deltabar q_{cd} + \tfrac12\,\mathcal D_a(\deltabar u_{2b}) + (a\leftrightarrow b)\,. \label{deltaXi}
\end{align}
Also for other objects we have:
\begin{align}
\delta\omega_a = &
\tfrac{1}{2} ( \eta^{b}- \overline{\eta}^{b} )\deltabar q_{ab}
+(\tfrac{1}{2} \Xi_{a}{}^b + \mathit{k}_{a} \omega^{b}+\tfrac{1}{2} \overline{\kappa}\delta^b_a) \deltabar u_{1b} 
+ (- \tfrac{1}{2} \Theta_{a}{}^b + \ell_{a} \omega^{b}+ \tfrac{1}{2}  \kappa\,\delta^b_a)\deltabar u_{2b}\nonumber\\
&	-  \tfrac{1}{2} \overline{\mathit{a}}_{a} \deltabar  \mu_{1} + \tfrac{1}{2} \mathit{a}_{a} \deltabar \mu_{2} + \tfrac{1}{2}  (- \overline{\eta}_{a} + \eta_{a})\deltabar \mu_{3} \nonumber\\
&	   -  \tfrac{1}{2}\,q^b{}_{a}\, \mathit{k}^{c} \nabla_{c}(\deltabar u_{1b}) + \tfrac{1}{2}\,q^b{}_{a}\, \ell^{c} \nabla_{c}(\deltabar u_{2b}) -  \tfrac{1}{2}\,\mathcal D_a(\deltabar \mu_{3}) - \mathcal D_a(\delta \ln\! D)\,. \label{deltaOmege}
\end{align}
\begin{align}
\delta a_a =  \mathit{k}_{a}\,\mathit{a}^{b}\, \deltabar u_{1b} + \ell_{a}\,\mathit{a}^{b} \, \deltabar u_{2b} + (\omega_a -  \tfrac{1}{2}\, \eta_{a} +\tfrac{1}{2}\,\overline{\eta}_{a}) \deltabar \mu_{1} -  \mathit{a}_{a} (\deltabar \mu_{3} + 2\, \delta \ln\! D)  -  \tfrac{1}{2} \mathcal D_a(\deltabar \mu_{1}) \,.\label{deltaa}
\end{align}
\begin{align}
\delta \bar a_a =  \mathit{k}_{a}\,\bar{\mathit{a}}^{b}\, \deltabar u_{1b} + \ell_{a}\,\bar{\mathit{a}}^{b} \, \deltabar u_{2b} - (\omega_a -  \tfrac{1}{2}\, \eta_{a} +\tfrac{1}{2}\,\overline{\eta}_{a}) \deltabar \mu_{2} +  \bar{\mathit{a}}_{a} (\deltabar \mu_{3} + 2\, \delta \ln\! D)  -  \tfrac{1}{2} \mathcal D_a(\deltabar \mu_{2})\,. \label{deltaba}
\end{align}
\begin{align}
\delta \kappa = (\omega^{a}-\eta^a)\deltabar u_{1a}  - \mathit{a}^{a} \deltabar u_{2a} -  \tfrac{1}{2} \overline{\kappa} \deltabar \mu_{1}  -   \kappa\,\delta \ln\! D  + \tfrac{1}{2} \mathit{k}^{a} \nabla_{a}(\deltabar \mu_{1}) -  \ell^{a} \nabla_{a}(\deltabar \mu_{3}) -  \ell^{a} \nabla_{a}(\delta \ln\! D)\,. \label{deltakappa}
\end{align}
\begin{align}
\delta \bar\kappa = \bar{\mathit{a}}^{a} \deltabar u_{1a} + (\omega^{a}+\bar\eta^a)\deltabar u_{2a}   -  \tfrac{1}{2} \,\kappa \deltabar \mu_{2}  +   \bar\kappa(\deltabar \mu_3+\delta \ln\! D)  - \tfrac{1}{2} \ell^{a} \nabla_{a}(\deltabar \mu_{2}) -  k^{a} \nabla_{a}(\delta \ln\! D)\,. \label{deltabk}
\end{align}

\section{Explicit Calculation of $\nabla_a\delta g_{bc}$}\label{Explicit}
\begin{align}
\nabla_a(\delta g_{bc}) =& \tfrac12\,\nabla_a(\deltabar q_{bc}) +k_b\,\nabla_a(\deltabar u_{1c}) + \ell_b\,\nabla_a(\deltabar u_{2c}) \nonumber\\
& + \tfrac12\,k_b\,k_c\,\nabla_a(\deltabar \mu_1) + \tfrac12\,\ell_b\,\ell_c\,\nabla_a(\deltabar \mu_2) + \ell_b\,k_c\,\nabla_a(\deltabar \mu_3) \nonumber\\
& +(\deltabar u_{1b} + k_b\,\deltabar \mu_1 + \ell_b\,\deltabar \mu_3)\nabla_ak_c +(\deltabar u_{2b} + \ell_b\,\deltabar \mu_2 + k_b\,\deltabar \mu_3)\nabla_a\ell_c \nonumber\\
&
+ (b\leftrightarrow c)
\end{align}

Decomposing this expression, using $\delta^a{}_b = q^a{}_b - \ell^a\,k_b - k^a\,\ell_b$, we get:
\begin{align}
\nabla_a(\delta g_{bc}) =&  \tfrac12\,\mathcal D_a(\deltabar q_{bc}) - \tfrac12\,\mathcal \ell_{a}\,q^e{}_{b}\, q^f_{}{c}\, \mathit{k}^{d}\,  \nabla_{d}(\deltabar q_{ef}) -  \tfrac12\,\mathit{k}_{a}\,q^e{}_{b}\, q^f{}_{c}\, \ell^{d} \nabla_{d}(\deltabar q_{ef})\nonumber \\ 
&+ \mathit{k}_{c}\,\mathcal D_a(\deltabar u_{1 b}) - \ell_{a}\,\mathit{k}_{b}\, q_{ce}\,\mathit{k}^{d}\, \nabla_{d}(\deltabar u_{1}{}^{e})  -  \mathit{k}_{a}\,\mathit{k}_{b}\,q_{ce}\,\ell^{d}\,\nabla_{d}(\deltabar u_{1}{}^{e})   \nonumber \\ 
& + \ell_{c} \mathcal{D}_a(\deltabar u_{2_b}) -  q_{ce} \mathit{k}^{d} \ell_{a} \ell_{b} \nabla_{d}(\deltabar u_{2}{}^{e}) -   \mathit{k}_{a}\,\ell_{b}\,q_{ce}\,\ell^{d}\,\nabla_{d}(\deltabar u_{2}{}^{e})  \nonumber \\ 
& + \tfrac12\mathit{k}_{b} \mathit{k}_{c} \mathcal D_a(\deltabar \mu_{1}) -\tfrac12\,\ell_{a}\,\mathit{k}_{b}\,\mathit{k}_{c}\,\mathit{k}^{d}\,  \nabla_{d}(\deltabar \mu_{1}) - \tfrac12\,\mathit{k}_{a}\,\mathit{k}_{b}\, \mathit{k}_{c}\,\ell^{d}\,\nabla_{d}(\deltabar \mu_{1}) \nonumber \\ 
&+ \tfrac12\ell_{b} \ell_{c} \mathcal D_a(\deltabar \mu_{2}) -\tfrac12\,\ell_{a}\,\ell_{b}\,\ell_{c}\,\mathit{k}^{d}\,\nabla_{d}(\deltabar \mu_{2})  - \tfrac12\,\mathit{k}_{a}\,\ell_{b}\,\ell_{c}\,\ell^{d}\,\nabla_{d}(\deltabar \mu_{2})\nonumber \\ 
& +\ell_{b}\,\mathit{k}_{c} \,\mathcal D_a(\deltabar \mu_{3}) -  \ell_{a}\, \ell_{b}\,\mathit{k}_{c}\,\mathit{k}^{d}\,  \nabla_{d}(\deltabar \mu_{3})  -  \mathit{k}_{a}\,\mathit{k}_{b}\,\ell_{c}\, \ell^{d}\,\nabla_{d}(\deltabar \mu_{3})  \nonumber \\ 
&  + (k_b\,\deltabar q_{cd} + k_b\,k_c\,\deltabar u_{1d} + g_{cd}\,\deltabar u_{2b} +\ell_b\,k_c\,\deltabar u_{2d} + g_{cd}\,\ell_b\,\deltabar \mu_2 + g_{cd}\,k_b\,\deltabar \mu_3)\nabla_a \ell^d \nonumber \\ 
&+ (\ell_b\,\deltabar q_{cd} + g_{cd}\,\deltabar u_{1b} + \ell_b\,\ell_c\,\deltabar u_{2d}  + \ell_b\,k_c\,\deltabar u_{1d} + g_{cd}\,k_b\,\deltabar \mu_1 + g_{cd}\,\ell_b\,\deltabar \mu_3)\nabla_a k^d \label{deldeltagnull2}
\end{align}
To derive the above expression we have used the following relations:
\begin{align}\label{BPrelations}
& \ell^b\,\nabla_a \deltabar q_{bc} = - \deltabar q_{bc}\,\nabla_a\ell^b\ ,\quad
\ell^b\,\nabla_a \deltabar u_{1b} = - \deltabar u_{1b}\,\nabla_a\ell^b \ ,\quad
\ell^b\,\nabla_a \deltabar u_{2b} = - \deltabar u_{2b}\,\nabla_a\ell^b \nonumber\\
& k^b\,\nabla_a \deltabar q_{bc} = - \deltabar q_{bc}\,\nabla_a k^b\ ,\quad
k^b\,\nabla_a \deltabar u_{1b} = - \deltabar u_{1b}\,\nabla_a k^b \ ,\quad
k^b\,\nabla_a \deltabar u_{2b} = - \deltabar u_{2b}\,\nabla_a k^b 
\end{align}

To write the expression \eqref{deldeltagnull2} in terms of variation of geometrical objects associated with the null hypersurface, namely $\delta \Theta_{ab}, \delta\Xi_{ab}, \delta\omega_a, \delta\eta_a, \delta \bar\eta , \delta\kappa$ and $\delta\bar{\kappa}$, we use the results of appendix \ref{appendix:variation of geometric objects}, the \cref{deltaTheta,deltaOmege,deltaXi,deltaa,deltaba,deltabk,deltakappa}, and find:
\begin{align}
\nabla_a(\delta g_{bc})=& -k_a\,\delta\Theta_{bc} -\ell_a\,\delta\Xi_{bc}+2\,\ell_a\,k_b\,\delta\omega_{c}-\delta a_a\,k_b\,k_c -\delta\bar a_a\,\ell_b\,\ell_c \nonumber\\
&
+\left (2\, \Theta_{(a}{}^{d} \mathit{k}_{b)}\, \delta_{c}{}^{e}  +2\, \Xi_{(a}{}^{d}\,\ell_{b)}\, \delta_{c}{}^{e} + 2\,  \ell_{(a}\,\mathit{k}_{b)}\,\delta_{c}{}^{d}\,\overline{\eta}^{e}  - \mathit{k}_{a}\, \mathit{k}_{b}\, \delta_{c}{}^{d}\, \mathit{a}^{e} -   \ell_{a}\, \ell_{b}\,\delta_{c}{}^{d}\,\overline{\mathit{a}}^{e}\right)\,\deltabar q_{de}\nonumber\\
&
+\left( \Theta_{a}{}^{d}\, \mathit{k}_{b}\, \mathit{k}_{c} + 2\, \mathit{k}_{a}\,\Theta^{d}{}_{b}\, \mathit{k}_{c}  +  \Xi_{ab}\,\delta_{c}{}^{d}+ 2\, \ell_{(a} \Xi_{b)}{}^{d}\,\mathit{k}_{c}   \right.\nonumber\\
& \left.\qquad
- 2\,  \omega_{(a}\, \mathit{k}_{b)}\,\delta_{c}{}^{d} - \ell_{a} \, \mathit{k}_{b}\, \mathit{k}_{c} \,\omega^{d}+ \mathit{k}_{a}\, \mathit{k}_{b}\, \ell_{c} \,\overline{\eta}^{d} -    \ell_{a}\,\mathit{k}_{b}\, \ell_{c}\,\overline{\mathit{a}}^{d} +  \kappa\,\mathit{k}_{a}\, \mathit{k}_{b}\,\delta_{c}{}^{d} \right)\deltabar u_{1d}\nonumber\\
&
+\left( \Theta_{ab}\,\delta_{c}{}^{d}  + 2\,\mathit{k}_{(a} \, \Theta_{b)}{}^{d}\,\ell_{c} + 2\, \ell_{(a} \,\mathit{k}_{b)}\,\Theta_{c}{}^{d} + \Xi_{a}{}^{d}\,\ell_{b}\, \ell_{c}  + 2\, \ell_{a}\,\Xi_{b}{}^{d}\, \ell_{c} + 2\,\omega_{(a}\,\ell_{b)}\,\delta_{c}{}^{d} \right.\nonumber\\
&
\left. \qquad   - 2\, \ell_{a}\, \ell_{b}\,\mathit{k}_{c}\, \omega^{d} -  \mathit{k}_{a}\, \ell_{b}\, \ell_{c}\, \omega^{d} + \ell_{a}\, \ell_{b}\,\mathit{k}_{c} \, \eta^{d}  - \mathit{k}_{a}\, \mathit{k}_{b}\, \ell_{c}\,\mathit{a}^{d} - 2\, \kappa\, \ell_{(a}\,\mathit{k}_{b)}\,\delta_{c}{}^{d}  - \overline{\kappa}\, \ell_{a}\, \ell_{b}\,\delta_{c}{}^{d}  \right)\deltabar u_{2d} \nonumber\\
&
+\left(\tfrac12\,\mathit{k}_ {a}\,\Xi_{bc}+  \Xi_{ab}\,\mathit{k}_ {c}  - \omega_{a}\,\mathit{k}_ {b}\,\mathit{k}_ {c}+ \mathit{k}_ {a}\, \mathit{k}_{b}\,\overline{\eta}_ {c} + \kappa\,\mathit{k}_ {a}\,\mathit{k}_ {b}\,\mathit{k}_ {c}   + \tfrac12\,\overline{\kappa}\,\ell_{a}\,\mathit{k}_ {b}\,\mathit{k}_ {c} \right)\deltabar \mu_{1} \nonumber\\
& +  \left(\tfrac12\, \ell_{a}\, \Theta_{bc}+  \Theta_{ab}\,\ell_{c} +  \omega_{a}\,\ell_{b}\, \ell_{c} + \ell_{a} \ell_{b} \eta_{c} -  2\,\ell_{(a}\,\mathit{k}_ {b)}\,\mathit{a}_ {c} -  \tfrac12\,\kappa\,\mathit{k}_ {a}\, \ell_{b}\, \ell_{c}  - \overline{\kappa}\, \ell_{a}\, \ell_{b}\, \ell_{c}    \right)\deltabar \mu_{2}\nonumber\\
& + \tfrac12\,\mathcal D_a(\deltabar q_{bc}) + 2\,k_{(a}\,\mathcal D_{b)}(\deltabar u_{1c}) - k_a\,k_b\,q^e{}_{c}\,\ell^d\,\nabla_d(\deltabar u_{1e}) + 2\,\ell_{(a}\,\mathcal D_{b)}(\deltabar u_{2c}) -2\, \ell_a\,\ell_{(b}\,k_{d)}\,q^e{}_{c}\,\nabla^d(\deltabar u_{1e})\nonumber\\
& +\tfrac12\,\mathcal D_a(\deltabar \mu_1)\,k_b\,k_c - \tfrac12\,k_a\,k_b\,k_c\ell^d\,\nabla_d(\deltabar \mu_1) + \tfrac12\,\mathcal D_a(\deltabar \mu_2)\,\ell_b\,\ell_c - \tfrac12\,\ell_a\,\ell_b\,\ell_c\,k^d\,\nabla_d(\deltabar \mu_2)\nonumber\\
& +2\,\ell_{(a}\,\mathcal D_{b)}(\deltabar \mu_3) \,k_c - \ell_a\,\ell_b\,k_c\,k^d\,\nabla_d(\deltabar \mu_3)-2\,\ell_{(a}\,k_{b)}\,k_c\,k^d\,\nabla_d(\deltabar \mu_3) \nonumber\\
&+ 2\,\ell_a\,k_b\,\mathcal D_c(\delta \ln\! D) - \ell_a\,k_b\,k_c\,\ell^d\,\nabla_d(\delta \ln\! D) + k_a\,\ell_b\,\ell_c\,k^d\,\nabla_d(\delta \ln\! D)\nonumber\\
&+ (b\leftrightarrow c) \,.
\end{align}

\section{Complete Decomposition of the Riemann tensor}\label{app:decompositionEOM}
In this appendix, we first decompose the Riemann tensor in normal and tangential to the co-dimension two surfaces $\mathcal S$. These are counterparts to the Gauss-Coddazi-Ricci equations in standard 3+1 formalism. The components of spacetime curvature tensor are expressed in terms of intrinsic and extrinsic geometrical objects of the co-dimension two surfaces $\mathcal S$. The number of independent components of the Riemann tensor in d-dimensions is $\tfrac1{12} d^2(d^2-1)$. By contracting the indices of the Riemann tensor with the projector $q^a{}_b$ and vectors $\ell^a$ and $k^a$, we can find all of structures and their number in table \ref{countingtable}. There we have used the symmetries of the Riemann tensor in counting the independent components of each structure which sum up to match that of the Riemann tensor. Among these symmetries are the Bianchi identities, which contribute in counting, by a subtraction of table $({}^{d-2}_{\ \,3})$
in second line of table \ref{countingtable}.

\begin{table}\caption{Independent components of the Riemann tensor in $2+(d-2) $ decomposion}\label{countingtable}
	\begin{center}
		\begin{tabular}{c|c}
			structure & number of independent components \\ \hline \\
			$qqqqR $&$\qquad \tfrac1{12}(d-2)^2((d-2)^2-1)$\\ \\
			$qqq\ell R , qqqk R$ &$\qquad 2\times\left[\tfrac12(d-2)^2(d-3) - \left({}^{d-2}_{\ \,3}\right)\right]$ \\ \\
			$q\ell q \ell R\,,\, q k q k R $&$\qquad 2\times \tfrac12(d-1)(d-2) $\\ \\
			$q\ell q k R $&$\qquad (d-2)^2 $\\ \\
			$q\ell\ell k R\,,\, qkk\ell R $&$\qquad 2\times(d-2)$ \\ \\
			$\ell k \ell k R $&$\qquad 1$ \\ 
		\end{tabular}
	\end{center}
\end{table}

In the following, we compute each of the structures \ref{countingtable} in terms of geometrical properties of the co-dimension two surface $\mathcal S$. The Gauss relation can be obtained by using Ricci equation on $\mathcal S$ :
\begin{equation}
[\mathcal{D}_a,\mathcal{D}_b]X^c=-\mathcal{R}_{abd}{}^c X^d
\end{equation}
where $ X $ is an arbitrary vector tangent to $ \mathcal{S} $, and $ \mathcal{D} $ is the covariant derivative compatible with its induced metric  and $ \mathcal{R} $ is the intrinsic curvature of that surface. Using the definition of $ \mathcal{D} $: $ D_aV_b=q_a^cq_b^d\nabla_cV_d $, and the relation for $ q $, we can find:
\begin{equation}
\mathcal{R}_{abd}{}^c X^d=q^{ce} q_{a}{}^{f} q_{b}{}^{g} R_{defg} X^{d} +  \Theta_{bd}\, \Xi_{a}{}^{c}\,X^{d} -   \Theta_{b}{}^{c}\, \Xi_{ad}\,X^{d}
-   \Theta_{ad}\, \Xi_{b}{}^{c}\,X^{d} +  \Theta_{a}{}^{c}\, \Xi_{bd}\,X^{d}
\end{equation}
Because $ X $ is general, we can write the above relation as 
\begin{equation}
q_{a}{}^{e} q_{b}{}^{f} q_{c}{}^{g} q_{d}{}^{h}R_{efgh}  =\mathcal{R}_{abcd} +  \Theta_{bd}\, \Xi_{a}{}^{c} -   \Theta_{b}{}^{c}\, \Xi_{ad}
-   \Theta_{ad}\, \Xi_{b}{}^{c} +  \Theta_{a}{}^{c}\, \Xi_{bd}\label{GC1relation}
\end{equation}
in which LHS is of the first structure in table \ref{countingtable}.

To obtain other relations, we can use the Ricci relation in spacetime:
\begin{equation}
[\nabla_a,\nabla_b]V^c=-R_{abd}{}^cV^d \label{riccirealtion}
\end{equation}
where $ V $ is an arbitrary vector on spacetime. Substituting one of $ \ell $ or $ k $ instead of $ V $ and contracting the indices in both side of \eqref{riccirealtion}, with $ q $, $ \ell $ or $ k $, we can find all other relations as follows:
\begin{align}
q_{a}{}^{e} q_{b}{}^{f} q_{c}{}^{g} R_{efgd} \ell^{d}&=- \Theta_{bc} \omega_{a} + \Theta_{ac} \omega_{b} +  \mathcal{D}_a\Theta_{bc} -  \mathcal{D}_b\Theta_{ac}\label{GC2relation}\\
\nonumber\\
q_{a}{}^{e} q_{b}{}^{f} q_{c}{}^{g} R_{efgd} k^{d}&=\Xi_{bc} \omega_{a} -  \Xi_{ac} \omega_{b} + \mathcal{D}_a\Xi_{bc} -  \mathcal{D}_b\Xi_{ac}\label{GC3relation}\\
\nonumber\\
q_{a}{}^{e} \ell^{f} q_{c}{}^{g} R_{efgd} \ell^{d}&=- \overline{\eta}_{a} \mathit{a}_{c} -  \mathit{a}_{a} \eta_{c} -  \Theta_{a}{}^{e} \Theta_{ce} + \Theta_{ac} \kappa - 2 \mathit{a}_{c} \omega_{a} -  q_{a}{}^{f} q_{c}{}^{g} \ell^{e} \nabla_{e}\Theta_{fg} + \mathcal{D}_a\mathit{a}_c
\nonumber\\
&=- \overline{\eta}_{a} \mathit{a}_{c} -  \mathit{a}_{a} \eta_{c} +  \Theta_{a}{}^{e} \Theta_{ce} + \Theta_{ac} \kappa - 2 \mathit{a}_{c} \omega_{a} -  q^* \mathscr L_\ell\Theta_{ac} + \mathcal{D}_a\mathit{a}_c \label{GC4relation}
\\\nonumber\\
q_{a}{}^{e} k^{f} q_{c}{}^{g} R_{efgd} k^{d}&=- \overline{\mathit{a}}_{a} \overline{\eta}_{c} -  \overline{\mathit{a}}_{c} \eta_{a}-  \Xi_{a}{}^{e} \Xi_{ce} -  \overline{\kappa} \Xi_{ac}  + 2 \overline{\mathit{a}}_{c} \omega_{a} -  q_{a}{}^{f} q_{c}{}^{g} \mathit{k}^{e} \nabla_{e}\Xi_{fg} + \mathcal{D}_a\overline{\mathit{a}}_{c}\nonumber\\
&=- \overline{\mathit{a}}_{a} \overline{\eta}_{c} -  \overline{\mathit{a}}_{c} \eta_{a} + \Xi_{a}{}^{e} \Xi_{ce} -  \overline{\kappa} \Xi_{ac}  + 2 \overline{\mathit{a}}_{c} \omega_{a} -  q^* \mathscr L_k\Xi_{ac} + \mathcal{D}_a\overline{\mathit{a}}_{c}\label{GC5relation}\\
\nonumber\\
q_{a}{}^{e} \ell^{f} q_{c}{}^{g} R_{efgd} k^{d}&=\overline{\mathit{a}}_{c} \mathit{a}_{a} + \eta_{a} \eta_{c} + \overline{\kappa} \Theta_{ac} -  \Theta_{a}{}^{d} \Xi_{cd} -  q_{a}{}^{e} q_{c}{}^{g} \mathit{k}^{d} \nabla_{d}\Theta_{eg} -\mathcal{D}_c\eta_{a}\nonumber\\
&=\overline{\mathit{a}}_{c} \mathit{a}_{a} + \eta_{a} \eta_{c} + \overline{\kappa} \Theta_{ac} +  \Theta_{c}{}^{d} \Xi_{ad} -  q^* \mathscr L_k\Theta_{ac} -\mathcal{D}_c\eta_{a}\label{GC6relation}\\
\nonumber\\
q_{a}{}^{e} \ell^{f} k^{g} R_{efgd} \ell^{d}&=- \overline{\kappa} \mathit{a}_{a} -  \overline{\eta}^{e} \Theta_{ae} + \overline{\eta}_{a} \kappa -  \mathit{a}^{e} \Xi_{ae} + \kappa \omega_{a} + \Theta_{ae} \omega^{e} + q_{af} \ell^{e} \nabla_{e}\omega^{f} -  \mathcal{D}_a\kappa\nonumber\\
&=- \overline{\kappa} \mathit{a}_{a} -  \overline{\eta}^{e} \Theta_{ae} + \overline{\eta}_{a} \kappa -  \mathit{a}^{e} \Xi_{ae} + \kappa \omega_{a} + q^*\mathscr L_\ell\omega^{a} -  \mathcal{D}_a\kappa\label{GC7relation}\\
\nonumber\\
q_{a}{}^{e} k^{f} \ell^{g} R_{efgd} k^{d}&=- \overline{\kappa} \eta_{a} -  \overline{\mathit{a}}^{d} \Theta_{ad} + \overline{\mathit{a}}_{a} \kappa -  \eta^{d} \Xi_{ad} + \overline{\kappa} \omega_{a} -  \Xi_a^d \omega_{d} -  q_{ae} \mathit{k}^{d} \nabla_{d}\omega^{e} + \mathcal{D}_a\overline{\kappa}\nonumber\\
&=- \overline{\kappa} \eta_{a} -  \overline{\mathit{a}}^{d} \Theta_{ad} + \overline{\mathit{a}}_{a} \kappa -  \eta^{d} \Xi_{ad} + \overline{\kappa} \omega_{a}  -  q^*\mathscr L_k\omega^{e} + \mathcal{D}_a\overline{\kappa}\label{GC8relation}\\
\nonumber\\
\ell^{e} k^{f} \ell^{g} R_{efgd} k^{d}&=- \overline{\mathit{a}}^{d} \mathit{a}_{d} + \overline{\eta}^{d} \eta_{d} + 2 \overline{\kappa} \kappa + \overline{\eta}^{d} \omega_{d} -  \eta^{d} \omega_{d} + \mathscr L_\ell\overline{\kappa} -  \mathscr L_k\kappa\label{GC9relation}
\end{align}
In these equations, $q^*\mathscr L$ is the Lie derivative projected on $\mathcal S$, e.g. $q^*\mathscr L_\ell \Theta_{ab}= q_{a}{}^c\,q_b{}^d\,\mathscr L_\ell\Theta_{cd}$. \Cref{GC1relation,GC2relation,GC3relation,GC4relation,GC5relation,GC6relation,GC7relation,GC8relation,GC9relation} give the complete set of nine different structures of contracted Riemann tensor in table \ref{countingtable}.

Similar relations can be obtained for the Ricci tensor as follows:
\begin{align}
 q_{c}{}^{a} q_{d}{}^{b} R_{ab} &=g^{ef}\,q_{c}{}^{a}\, q_{d}{}^{b}\, R_{eafb} 
 =(q^{ef}-\ell^e k^f - k^e\ell^f)\,q_{c}{}^{a}\, q_{d}{}^{b}\, R_{eafb} \nonumber\\
 &= \mathcal R_{cd} - 4\Theta^a{}_{(c}\,\Xi_{d)a} - \Theta_{cd} \Xi^{a}{}_{a} - \Theta^{a}{}_{a} \Xi_{cd}  - 2 \overline{\kappa} \Theta_{cd}  - 2 \eta_{c} \eta_{d} -  2\,a_{(c}\overline{a}_{d)}\label{qqR}    \nonumber\\
 &\quad+ 2\, q^*\mathscr L_k \Theta_{cd} + 2\,\mathcal D_{(c}\eta_{d)}\\
 \nonumber\\
 q_a{}^c\,\ell^b\,R_{bc} &= \Theta\, \omega_{a} -\Theta_{ab}\,\omega^b + \eta^{c} \Theta_{ac} + \overline{\eta}_{a} \kappa -  \mathit{a}^{c} \Xi_{ac}  + \kappa \omega_{a} - \overline{\kappa} \mathit{a}_{a}\nonumber\\
 &\quad + q^*\mathscr L_\ell \omega_a + \mathcal D_b\Theta_{a}{}^{b} -  \mathcal D_a\Theta^{b}{}_{b}  -  \mathcal D_a \kappa \label{qlR}\\
 \nonumber\\
 q_a{}^c\,k^b\,R_{bc} &=-  \Xi\, \omega_{a} -\Xi_{ab}\,\omega^b - \overline{\kappa} \eta_{a} -  \overline{\mathit{a}}^{c} \Theta_{ac} + \overline{\mathit{a}}_{a} \kappa + \overline{\eta}^{c} \Xi_{ac} + \overline{\kappa} \omega_{a}\nonumber\\
 &\quad   -  q^*\mathscr L_k\omega_a + \mathcal D_b\Xi_{a}{}^{b} -  \mathcal D_a\Xi + \mathcal D_a\overline{\kappa}\label{qkR}\\
 \nonumber\\
 \ell^a\,\ell^b\,R_{ab} &= - \Theta_{ab}\, \Theta^{ab} + \Theta\, \kappa - 2 \mathit{a}^{a} \omega_{a} -a^a\,\eta_a - a^a\,\bar{\eta}_a -  \mathscr L_\ell\Theta + \mathcal D_{a}\mathit{a}^{a} \label{llR}\\
 \nonumber\\
 k^a\,k^b\,R_{ab} &=  -  \Xi_{ab}\, \Xi^{ab} - \bar\kappa \,\Xi + 2 \overline{\mathit{a}}^{a} \omega_{a} -\bar a^a\,\eta_a - \bar a^a\,\bar{\eta}_a -  \mathscr L_k \Xi + \mathcal D_a \bar a^{a} \label{kkR}\\
 \nonumber\\
 \ell^a\,k^b\,R_{ab} &= -  \Theta^{ab} \Xi_{ab} + \bar{\kappa} (\Theta + 2 \kappa)  + \overline{\eta}^{a} \omega_{a} -  \eta^{a} \omega_{a} + \eta^a\,\eta_a +\eta^a\,\bar{\eta}_a   \nonumber\\
 &\quad -  \mathscr L_k(\Theta+\kappa) + \mathscr L_{\ell}\bar{\kappa} -  \mathcal D_{b}\eta^{b}\label{lkR}
\end{align}

The Ricci scalar turns out to be
\begin{align}
R &= \mathcal R - 4\, \bar{\kappa} (\Theta +\kappa)  + 2\, \Theta^{ab} \,\Xi_{ab} - 2\, \Theta\, \Xi - 2\, \overline{\eta}^{a} \omega_{a} + 2\, \eta^{a} \omega_{a}  - 2 \overline{\eta}^{a} \eta_{a} - 4\, \eta_{a} \eta^{a} - 2\, \overline{\mathit{a}}^{a} \mathit{a}_{a}\nonumber\\
&\quad - 2\, \mathcal L_\ell\bar{\kappa} + 2\, \mathscr L_k(2\,\Theta+\kappa) + 4\, \mathcal D_a\eta^{a}\,.
\end{align}

Finally, we decompose the Einstein tensor. Projecting $G_{ab}$ on co-dimension two surfaces $\mathcal S$ gives
\begin{align}
q_{a}^c\,q_b^d\,G_{cd} &= 
\mathcal G_{ab}
-  \Theta_{ab} \,\Xi -  \Theta\, \Xi_{ab} + q_{ab} \Theta\, \Xi-  q_{ab} \Theta^{cd} \Xi_{cd} -4\, \Theta_{(a}{}^c\,\Xi_{b)c}
\nonumber\\
& - 2 \eta_{a} \eta_{b}+q_{ab}\,( 2 \, \eta_{c} \eta^{c} + \overline{\eta}^{c} \, \omega_{c} -   \eta^{c} \,\omega_{c}  + \overline{\eta}^{c}\, \eta_{c})
\nonumber\\
& - 2 \overline{\kappa} \Theta_{ab} + 2 \overline{\kappa} q_{ab} (\Theta  + \kappa) - 2\, \overline{\mathit{a}}_{(a} \mathit{a}_{b)}  +q_{ab}\, \overline{\mathit{a}}^{c} \, \mathit{a}_{c}         
\nonumber\\
& + 2\,q^* \mathscr L_k\Theta_{ab}  + q_{ab} \mathscr L_\ell\overline{\kappa}  - q_{ab} \mathscr L_k(2\,\Theta+\kappa)  + 2\,\mathcal D_{(a}\eta_{b)} - 2\, q_{ab}\, \mathcal D_c\eta^{c}\,,
\end{align}
where $\mathcal G_{ab}$ is the Einstein tensor on $\mathcal S$. It is identically zero in a four-dimensional spacetime, where $\mathcal S$ has two dimensions. The $q\ell G$, $qk G$, $\ell \ell G$ and $k k G$ components are the same as \cref{qlR,qkR,llR,kkR} for corresponding components of Ricci tensor. The last component $\ell kG$ is
\begin{align}
\ell^a\,k^b\,G_{ab} &=  \tfrac{1}{2} \mathcal{R} -  \Theta\, \Xi -  \overline{\kappa} \Theta -  \eta_{a} \eta^{a} - \overline{\mathit{a}}^{a} \mathit{a}_{a}     + \mathscr L_k \Theta + \mathcal D_a\eta^{a}\,.
\end{align}

\singlespacing
\bibliographystyle{hunsrt}
\bibliography{BTrefrences1}

\begin{thebibliography}{10}

\bibitem{Gibbons:1976ue}
G.~W. Gibbons and S.~W. Hawking.
\newblock {Action Integrals and Partition Functions in Quantum Gravity}.
\newblock {\em Phys. Rev.}, D15:2752--2756, 1977.

\bibitem{Parattu:2015gga}
K.~Parattu, S.~Chakraborty, B.~R. Majhi, and T.~Padmanabhan.
\newblock {A Boundary Term for the Gravitational Action with Null Boundaries}.
\newblock {\em Gen. Rel. Grav.}, 48(7):94, 2016, 1501.01053.

\bibitem{Lehner2016}
L.~Lehner, R.~C. Myers, E.~Poisson, and R.~D. Sorkin.
\newblock {Gravitational action with null boundaries}.
\newblock {\em Physical Review D}, 94(8):1--45, 2016, 1609.00207.

\bibitem{Hopfmuller2017}
F.~Hopfm{\"{u}}ller and L.~Freidel.
\newblock {Gravity degrees of freedom on a null surface}.
\newblock {\em Physical Review D}, 95(10):1--28, 2017, 1611.03096.

\bibitem{Jubb:2016qzt}
I.~Jubb, J.~Samuel, R.~Sorkin, and S.~Surya.
\newblock {Boundary and Corner Terms in the Action for General Relativity}.
\newblock {\em Class. Quant. Grav.}, 34(6):065006, 2017, 1612.00149.

\bibitem{Neiman:2012fx}
Y.~Neiman.
\newblock {On-shell actions with lightlike boundary data}.
\newblock 2012, hep-th/1212.2922.

\bibitem{Padmanabhan2014}
T.~Padmanabhan.
\newblock {A short note on the boundary term for the Hilbert action}.
\newblock {\em Modern Physics Letters A}, 29(08):1450037, 2014.

\bibitem{Myers1987}
R.~C. Myers.
\newblock {Higher-derivative gravity, surface terms, and string theory}.
\newblock {\em Physical Review D}, 36(2):392--396, 1987.

\bibitem{misner2017gravitation}
C.~Misner, K.~Thorne, J.~Wheeler, and D.~Kaiser.
\newblock {\em Gravitation}.
\newblock Princeton University Press, 2017.

\bibitem{Chakraborty2017}
S.~Chakraborty, K.~Parattu, and T.~Padmanabhan.
\newblock {A novel derivation of the boundary term for the action in
  Lanczos–Lovelock gravity}.
\newblock {\em General Relativity and Gravitation}, 49(9), 2017, 1703.00624.

\bibitem{Parattu2016}
K.~Parattu, S.~Chakraborty, and T.~Padmanabhan.
\newblock {Variational principle for gravity with null and non-null boundaries:
  a unified boundary counter-term}.
\newblock {\em European Physical Journal C}, 76(3):1--6, 2016, 1602.07546.

\bibitem{Sachs1962}
R.~K. Sachs.
\newblock On the characteristic initial value problem in gravitational theory.
\newblock {\em Journal of Mathematical Physics}, 3(5):908--914, 1962,
  https://doi.org/10.1063/1.1724305.

\bibitem{Hayward1993}
S.~A. Hayward.
\newblock {Dual-null dynamics of the Einstein field}.
\newblock {\em Classical and Quantum Gravity}, 10(4):779--790, 1993.

\bibitem{Brady1995}
P.~R. Brady, S.~Droz, W.~Israel, and S.~M. Morsink.
\newblock {Covariant double-null dynamics: {\$}(2+2){\$}-splitting of the
  Einstein equations}.
\newblock 1995, 9510040.

\bibitem{Epp1995}
R.~J. Epp.
\newblock {The Symplectic Structure of General Relativity in the Double-Null
  (2+2) Formalism}.
\newblock (November), nov 1995, 9511060.

\bibitem{Vickers2011}
J.~A. Vickers.
\newblock {Double null hamiltonian dynamics and the gravitational degrees of
  freedom}.
\newblock pages 3411--3428, 2011.

\bibitem{Hayward:1993my}
G.~Hayward.
\newblock {Gravitational action for space-times with nonsmooth boundaries}.
\newblock {\em Phys. Rev.}, D47:3275--3280, 1993.

\bibitem{wald2010general}
R.~Wald.
\newblock {\em General Relativity}.
\newblock University of Chicago Press, 2010.

\bibitem{Gourgoulhon2006}
E.~Gourgoulhon and J.~L. Jaramillo.
\newblock {A 3 + 1 perspective on null hypersurfaces and isolated horizons}.
\newblock 423:159--294, 2006.

\bibitem{Jezierski:2003hh}
J.~Jezierski.
\newblock {Geometry of null hypersurfaces}.
\newblock In {\em {Proceedings, 7th Hungarian Relativity Workshop (RW 2003):
  Sarospatak, Hungary, August 10-15, 2003}}, pages 255--264, 2004,
  gr-qc/0405108.

\bibitem{Jafari2018b}
G.~Jafari and S.~Aghapour.
\newblock {in preparation}.

\bibitem{hayward2013black}
S.~Hayward.
\newblock {\em Black Holes: New Horizons}.
\newblock New horizons. World Scientific, 2013.

\bibitem{Andersson:2005gq}
L.~Andersson, M.~Mars, and W.~Simon.
\newblock {Local existence of dynamical and trapping horizons}.
\newblock {\em Phys. Rev. Lett.}, 95:111102, 2005, gr-qc/0506013.

\bibitem{Chakraborty:2018dvi}
S.~Chakraborty and K.~Parattu.
\newblock {Null Boundary Terms for Lanczos-Lovelock Gravity}.
\newblock 2018, 1806.08823.

\bibitem{Cano:2018ckq}
P.~A. Cano.
\newblock {Lovelock action with nonsmooth boundaries}.
\newblock {\em Phys. Rev.}, D97(10):104048, 2018, 1803.00172.

\bibitem{Iyer:1994ys}
V.~Iyer and R.~M. Wald.
\newblock {Some properties of Noether charge and a proposal for dynamical black
  hole entropy}.
\newblock {\em Phys. Rev.}, D50:846--864, 1994, gr-qc/9403028.

\bibitem{Wald:1999wa}
R.~M. Wald and A.~Zoupas.
\newblock {A General definition of 'conserved quantities' in general relativity
  and other theories of gravity}.
\newblock {\em Phys. Rev.}, D61:084027, 2000, gr-qc/9911095.

\bibitem{Hopfmuller:2018fni}
F.~Hopfmüller and L.~Freidel.
\newblock {Null Conservation Laws for Gravity}.
\newblock {\em Phys. Rev.}, D97(12):124029, 2018, 1802.06135.

\bibitem{Chandrasekaran:2018aop}
V.~Chandrasekaran, E.~E. Flanagan, and K.~Prabhu.
\newblock {Symmetries and charges of general relativity at null boundaries}.
\newblock 2018, 1807.11499.

\bibitem{MARTINGARCIA2008597}
J.~M. Martín-García.
\newblock xperm: fast index canonicalization for tensor computer algebra.
\newblock {\em Computer Physics Communications}, 179(8):597 -- 603, 2008.

\bibitem{Nutma:2013zea}
T.~Nutma.
\newblock {xTras : A field-theory inspired xAct package for mathematica}.
\newblock {\em Comput. Phys. Commun.}, 185:1719--1738, 2014, 1308.3493.

\end{thebibliography}

\end{document}